\documentclass[sn-mathphys-ay]{sn-jnl}
\usepackage[utf8]{inputenc}

\usepackage{amsmath,amssymb,amsfonts,amsthm,mathrsfs}

\usepackage[title]{appendix}
\usepackage{graphicx}
\usepackage{textcomp}
\usepackage{booktabs}
\usepackage{colortbl}
\usepackage[table]{xcolor}
\usepackage{pifont}
\usepackage{microtype}
\usepackage{xspace}
\usepackage[most]{tcolorbox}
\usepackage{enumitem}
\usepackage{subcaption}
\usepackage{wrapfig}
\usepackage{pgf,pgfmath,etoolbox}
\usepackage{multirow}
\usepackage{tabularx} 

\usepackage{colortbl}
\usepackage{xcolor}
\usepackage{pgf}
\usepackage{pgfmath}
\usepackage{etoolbox}

\definecolor{cellgreen1}{rgb}{0.93,1.0,0.93}
\definecolor{cellgreen2}{rgb}{0.75,0.93,0.75}
\definecolor{cellgreen3}{rgb}{0.55,0.80,0.55}
\definecolor{cellgreen4}{rgb}{0.35,0.65,0.35}
\definecolor{cellgreen5}{rgb}{0.2,0.5,0.2}

\definecolor{cellblue1}{rgb}{0.9,0.95,1}
\definecolor{cellblue2}{rgb}{0.6,0.8,1}
\definecolor{cellblue3}{rgb}{0.4,0.7,1}
\definecolor{cellblue4}{rgb}{0.2,0.6,1}
\definecolor{cellblue5}{rgb}{0.0,0.4,0.9}

\newcommand{\scorecellgreen}[1]{%
    \pgfmathsetmacro{\value}{#1}%
    \ifboolexpr{ test {\ifdimcomp{\value pt}{>}{65pt}} }
        {\cellcolor{cellgreen5}\textbf{#1}}
        {%
        \ifboolexpr{ test {\ifdimcomp{\value pt}{>}{50pt}} }
            {\cellcolor{cellgreen4}#1}
            {%
            \ifboolexpr{ test {\ifdimcomp{\value pt}{>}{30pt}} }
                {\cellcolor{cellgreen3}#1}
                {%
                \ifboolexpr{ test {\ifdimcomp{\value pt}{>}{20pt}} }
                    {\cellcolor{cellgreen2}#1}
                    {\cellcolor{cellgreen1}#1}%
                }%
            }%
        }%
    }%

\newcommand{\scorecellblue}[1]{%
    \pgfmathsetmacro{\value}{#1}%
    \ifboolexpr{ test {\ifdimcomp{\value pt}{>}{60pt}} }
        {\cellcolor{cellblue5}\textbf{#1}}
        {%
        \ifboolexpr{ test {\ifdimcomp{\value pt}{>}{50pt}} }
            {\cellcolor{cellblue4}#1}
            {%
            \ifboolexpr{ test {\ifdimcomp{\value pt}{>}{30pt}} }
                {\cellcolor{cellblue3}#1}
                {%
                \ifboolexpr{ test {\ifdimcomp{\value pt}{>}{20pt}} }
                    {\cellcolor{cellblue2}#1}
                    {\cellcolor{cellblue1}#1}%
                }%
            }%
        }%
    }%

\definecolor{cellred1}{rgb}{1.0,0.93,0.93}
\definecolor{cellred2}{rgb}{1.0,0.8,0.8}
\definecolor{cellred3}{rgb}{1.0,0.6,0.6}
\definecolor{cellred4}{rgb}{1.0,0.4,0.4}
\definecolor{cellred5}{rgb}{0.9,0.2,0.2}

\newcommand{\boxmargin}{4pt}

\newtcolorbox{RQbox}{
    colback=blue!10!white,  
    arc = 0pt, outer arc = 0pt,
    boxsep=0pt, left = 3pt, right = 0pt, top = 0pt, bottom = 0pt, 
    leftrule=3pt, bottomrule=0pt, toprule=0pt, rightrule=0pt,
    left = \boxmargin, right = \boxmargin, top = \boxmargin, bottom = \boxmargin
}

\definecolor{c1}{cmyk}{0,0.6175,0.8848,0.1490}
\definecolor{c2}{cmyk}{0.1127,0.6690,0,0.4431}
\definecolor{c3}{cmyk}{0.3081,0,0.7209,0.3255}
\definecolor{c4}{cmyk}{0.6765,0.2017,0,0.0667}
\definecolor{c5}{cmyk}{0,0.8765,0.7099,0.3647}

\definecolor{lightgrey}{rgb}{0.93,0.93,0.93}

\newtcbox{\hlprimarytab}{on line, rounded corners, box align=base, colback=c3!10,colframe=white,size=fbox,arc=3pt, before upper=\strut, top=-2pt, bottom=-4pt, left=-2pt, right=-2pt, boxrule=0pt}
\newtcbox{\hlsecondarytab}{on line, box align=base, colback=red!10,colframe=white,size=fbox,arc=3pt, before upper=\strut, top=-2pt, bottom=-4pt, left=-2pt, right=-2pt, boxrule=0pt}

\definecolor{codegreen}{rgb}{0,0.6,0}
\definecolor{codegray}{rgb}{0.5,0.5,0.5}
\definecolor{codepurple}{rgb}{0.58,0,0.82}
\definecolor{backcolour}{rgb}{0.95,0.95,0.92}
\def\bench{\textsc{CodeIF-Bench}\xspace}

\theoremstyle{thmstyleone}%
\theoremstyle{thmstyletwo}%

\theoremstyle{thmstylethree}%

\usepackage{subcaption}

\begin{document}

\title{\bench: Evaluating Instruction-Following Capabilities of Large Language Models in Interactive Code Generation}

\author[1]{\fnm{Peiding} \sur{Wang}}\email{wangpeiding@buaa.edu.cn}
\author[1]{\fnm{Li} \sur{Zhang}}\email{lily@buaa.edu.cn}
\author*[1]{\fnm{Fang} \sur{Liu}}\email{fangliu@buaa.edu.cn}
\author*[2]{\fnm{Lin} \sur{Shi}}\email{shilin@buaa.edu.cn}
\author[1]{\fnm{Minxiao} \sur{Li}}\email{liminxiao@buaa.edu.cn}
\author[3]{\fnm{Bo} \sur{Shen}}\email{shenbo21@huawei.com}
\author[3]{\fnm{An} \sur{Fu}}\email{fuan1@huawei.com}
\affil[1]{School of Computer Science \& Engineering, State Key Laboratory of Complex \& Critical Software Environment, Beihang University}
\affil[2]{School of Software, Beihang University}
\affil[3]{Huawei Cloud Computing Technologies Co., Ltd. China}






\abstract{Large Language Models (LLMs) have demonstrated exceptional performance in code generation tasks and have become indispensable programming assistants for developers. However, existing code generation benchmarks primarily assess the functional correctness of code generated by LLMs in single-turn interactions. They offer limited insight into LLMs' abilities to generate code that strictly follows users' instructions in multi-turn interaction scenarios. In this paper, we introduce \bench, a benchmark for evaluating the instruction-following capabilities of LLMs in interactive code generation. Specifically, \bench incorporates nine types of verifiable instructions aligned with the real-world software development requirements, which can be independently and objectively validated through specified test cases, facilitating the evaluation of instruction-following capability in multi-turn interactions. In both \textit{Static Conversation} and \textit{Dynamic Conversation} settings, we evaluate the performance of 6 state-of-the-art LLMs and summarize the important factors, additional repository context and gradually increasing interaction history influencing the instruction-following ability of LLMs in multi-turn interactions. 
Furthermore, we identify the potential direction for improvement: context management.
The code and data are available at \href{https://github.com/zhu-zhu-ding/CodeIF-Bench}{https://github.com/zhu-zhu-ding/CodeIF-Bench}.
}




\maketitle

\section{Introduction}\label{sec1}

%
In recent years, the remarkable advancements in Large Language Models (LLMs) for code generation~\citep{zhu2024deepseek,roziere2023code,luo2023wizardcoder} have significantly enhanced developers' coding efficiency through an interactive and conversational paradigm~\citep{GitHub-Copilot,Cursor}. 
 
\begin{wrapfigure}{hr}{0.5\linewidth} 
    \centering
    \vspace{-2em} 
    \includegraphics[width=\linewidth]{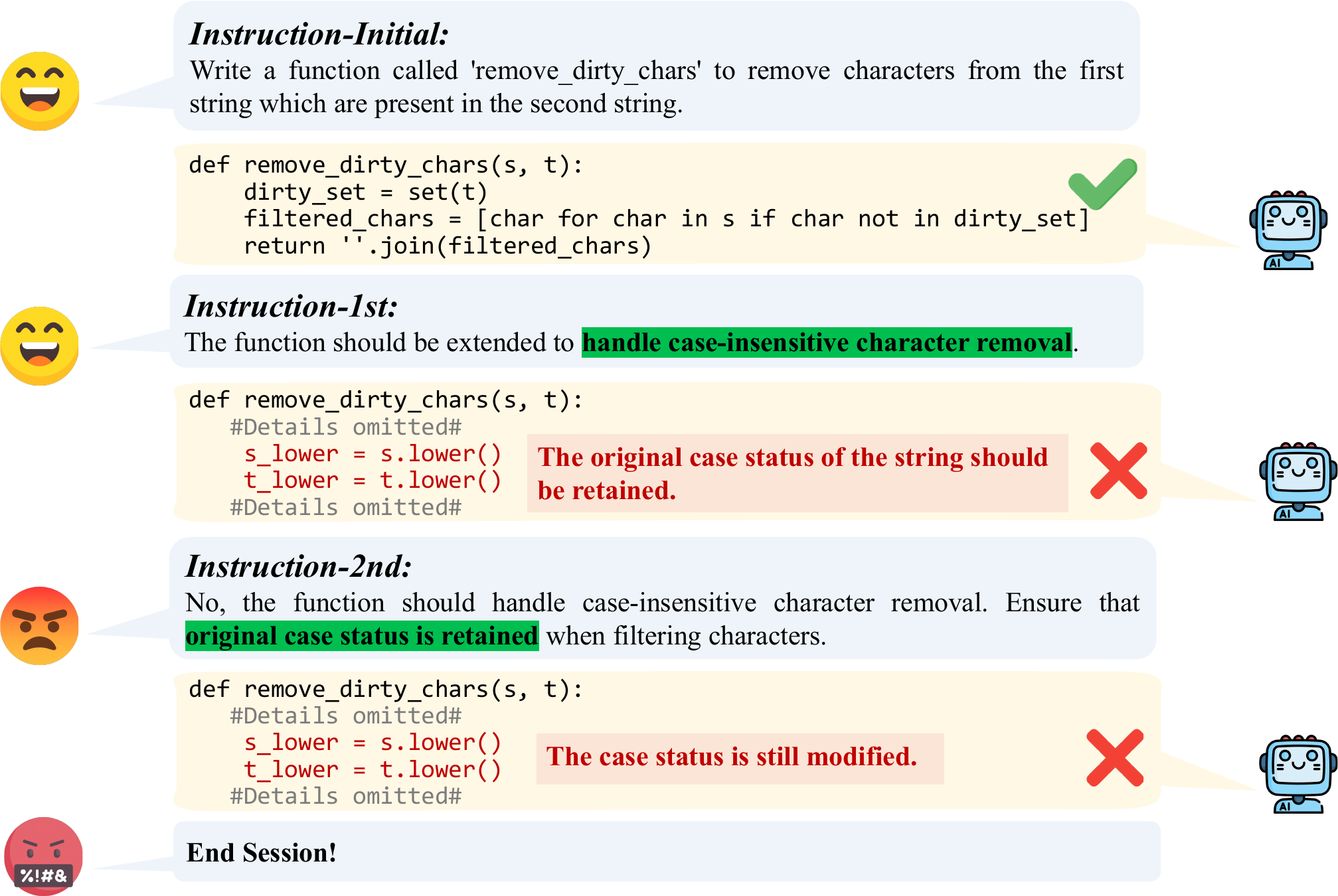}
    \caption{An example of interactive code generation, where the developer provides follow-up instructions to clarify the requirement and address issues in the generated code}
    \label{fig:Introduction_Example}
    \vspace{-2em} 
\end{wrapfigure}
In practical software development environments, as shown in Figure \ref{fig:Introduction_Example}, the collaboration between developers and LLMs typically necessitates an interactive code generation process. Throughout this interactive workflow, developers frequently enhance and refine their initial instructions by adding supplementary details to better align with their requirements. Specifically, through systematic code execution and rigorous review processes, developers often identify and provide follow-up instructions to clarify requirements and address issues in LLM-generated code. These issues may include implementing new features \citep{jimenez2023swe-bench}, fixing bugs \citep{codebugs}, resolving potential security vulnerabilities \citep{perry2023users}, \textit{etc.} Specifically, any discrepancies or misunderstandings in LLM following instructions can significantly increase interaction time and cost, as developers may need to spend additional time clarifying and fixing errors.  Besides, such misunderstandings might even trigger uncontrollable and unexpected issues, such as bugs \citep{codebugs}, system failures \citep{codeweaknesses}, or even security vulnerabilities \citep{perry2023users}, which can have severe implications for the software's performance and developers' trust. Consequently, a dedicated benchmark to evaluate LLMs' \textbf{Instruction-Following} (\textbf{IF}) capabilities in interactive code generation is essential.

The key challenges in benchmarking instruction following for interactive code generation lie in capturing the iterative, linguistic, and practical complexities of real-world developer–LLM collaboration:
\begin{itemize}[leftmargin=*]
    \item  \textbf{Challenge 1 - Modeling multi-turn interactions.} Unlike single-turn tasks, interactive code generation requires the benchmark to simulate iterative developer–LLM dialogues where each instruction depends on the previous interaction history. Constructing such multi-turn data is non-trivial, as it requires coherent context linking, and realistic instruction evolution. 
    \item  \textbf{Challenge 2 — Ambiguity and context dependence of instructions.} Developers’ instructions are often ambiguous, partial, or context-dependent. Accurately determining whether an LLM has truly ``followed'' such instructions is challenging.
    \item  \textbf{Challenge 3 — Ensuring real-world representativeness.}
    To accurately assess LLM's performance in real-world settings, the benchmark must reflect how real developers formulate and refine coding intents, rather than relying on generic natural language instructions that lack authentic development context and have already been extensively explored.
\end{itemize}
For \textbf{Challenge 1}, existing popular code generation benchmarks \citep{HumanEval,MBPP,APPS,DevEval} primarily evaluate whether LLMs can generate functionally correct code given specific programming tasks in a single turn, and it is difficult to directly evaluate LLMs' instruction-following capability in multi-round interaction. For \textbf{Challenge 2}, for instance, subjective instructions such as ``Please make the comments clearer'' may result in varying interpretations when assessing whether the model follows the given requirement. Some instruction-following benchmarks in the NLP domain, including for multi-turn \citep{multiif,MT-Bench, mt-bench-101}, most use the LLM-as-Judge method for evaluation, which may lead to bias \citep{li2025preferenceleakagecontaminationproblem,chen-etal-2024-humans}. 
To circumvent this challenge, some work, such as \citep{IF-Bench}, focuses on evaluating the ability of LLMs to follow ``verifiable instructions'' which are defined as instructions that can be objectively verified for compliance, \textit{e.g.}, ``the response should be more than 300 words''. However, it does not address the \textbf{Challenge 3}: these benchmarks do not align with real-world software development scenarios and fail to evaluate LLM's ability to follow instructions in realistic interactive coding tasks.

To address the above challenges, we propose a new benchmark named \bench, which aims to evaluate LLMs' instruction-following capability in interactive code generation. We first gather initial instructions from real-world projects \citep{DevEval} and crowd-sourced problems \citep{MBPP}, enabling evaluation for both standalone function and repository-level code generation. Then we use a combination of LLM and manual annotation to extract nine primary verifiable instructions strategies of real-world user requirements to guide LLM to generate \textbf{verifiable instructions},  which can be independently and objectively verified using their respective test cases, such as initial instructions. We guide LLM to generate verifiable instructions based on these strategies and initial instructions, resulting in each data point in \bench containing at least 8 verifiable instructions (8 for function-level and 10 for repository-level) to construct a conversation. In addition, we design \textit{Static Conversation} (predefined instruction sequences) and \textit{Dynamic Conversation} (providing feedback information), and evaluation metrics based on test cases to evaluate the performance of LLM in code interaction scenarios.

We evaluate the IF capabilities of 6 widely-used and advanced LLMs (\textit{i.e.} GPT-4o \citep{openai2024gpt},  Claude-3.5-Sonnet \citep{claude}, DeepSeek-V3 \citep{zhu2024deepseek}, Qwen2.5-Coder-\{7B, 14B, 32B\} \citep{qwencoder}) in interactive code generation tasks using \bench. Experimental results reveal that as additional repository contexts and dialogue histories are incorporated, the instruction-following performance of LLMs gradually declines. This degradation manifests as a weakened ability to follow the current instruction and an increasing tendency to forget previous instructions. In dynamic interactive settings with feedback, the growing contextual burden further hinders LLMs from effectively leveraging feedback to improve their outputs. Our case studies and analyses of various prompting strategies further suggest that effective context management constitutes a promising direction for enhancing LLM performance.

In summary, our contributions are as follows:
\begin{itemize}[leftmargin=*]
    \item
    To the best of our knowledge, we construct the first benchmark, \bench, for evaluating the instruction-following capabilities of LLMs in interactive code generation.
    \item 
    We present nine strategies derived from real software development to guide the verifiable instruction generation, which can be independently and objectively verified with their test cases.
    \item 
    We design \textit{Static Conversation} and \textit{Dynamic Conversation} interactive scenarios, and quantitative metrics based on test cases to evaluate the instruction following capabilities of LLMs.
    \item
    We perform a thorough evaluation of 6 prominent LLMs on \bench, analyzing the key factors influencing their IF capabilities in interactive code generation.
\end{itemize}

\section{Related Work}

\noindent \textbf{Code Generation Benchmarks.} 
Recent years have seen remarkable progress in LLM-based code generation, with models such as DeepSeek-Coder \citep{zhu2024deepseek}, WizardCoder \citep{luo2023wizardcoder}, and Qwen-Coder \citep{qwencoder} demonstrating exceptional performance. This progress has facilitated the development of interactive coding tools \citep{Cursor,GitHub-Copilot} that enhance programmer productivity through effective human-LLM collaboration. This development has driven the development of benchmarks in code generation. These include benchmarks for evaluating LLMs' performance on various code generation scenarios, including stand-alone function generation (\textit{e.g.}, HumanEval \citep{HumanEval}, MBPP \citep{MBPP}, APPS \citep{APPS}), class-level code generation (\textit{e.g.}, ClassEval \citep{ClassEval}), and repository-level code generation (\textit{e.g.}, DevEval \citep{DevEval} and CoderEval \citep{CoderEval}, which leverage real-world open-source repositories). While these benchmarks primarily assess whether LLMs can generate functionally correct code in single-turn interactions, they are insufficient for evaluating LLMs' interactive capabilities, particularly in multi-turn interaction scenarios.

\noindent \textbf{Instruction Following Benchmarks.}
Current benchmarks for evaluating the instruction-following capabilities of LLMs \citep{IF-Bench,MT-Eval,MT-Bench,FB-Bench} primarily focus on natural language QA tasks in the NLP domain. 
However, there is a significant disparity exists between code generation and QA tasks.
Some of the existing benchmarks utilize LLM-as-judge to assess LLMs' instruction following performance \citep{Systematic,huang2024empirical,chen2024mllm,MT-Bench}, which typically rely on elaborate prompts and can introduce inherent bias. CanItEdit \citep{CanItEdit} is a code editing benchmark, but its manually crafted programs lack project-specific dependencies, misaligning with real-world development. Additionally, its single-round dialogue design cannot assess LLMs' performance in complex, multi-turn interactions. CodeIF \citep{codeif} benchmarks instruction-following in code generation but oversimplifies real-world complexity. Its single-turn setting also fails to assess LLMs' multi-turn interaction capabilities.

\section{CodeIF-Bench}

In this section, we first give a definition of the interactive code generation (\textsection \ref{task_def}) task and the evaluation metric (\textsection \ref{metrics}).
Then we present the \bench construction pipeline (\textsection \ref{construction_pipeline}).

\begin{figure*}[h]
    \centering
    \setlength{\abovecaptionskip}{0.1cm}
    \includegraphics[width=\linewidth]{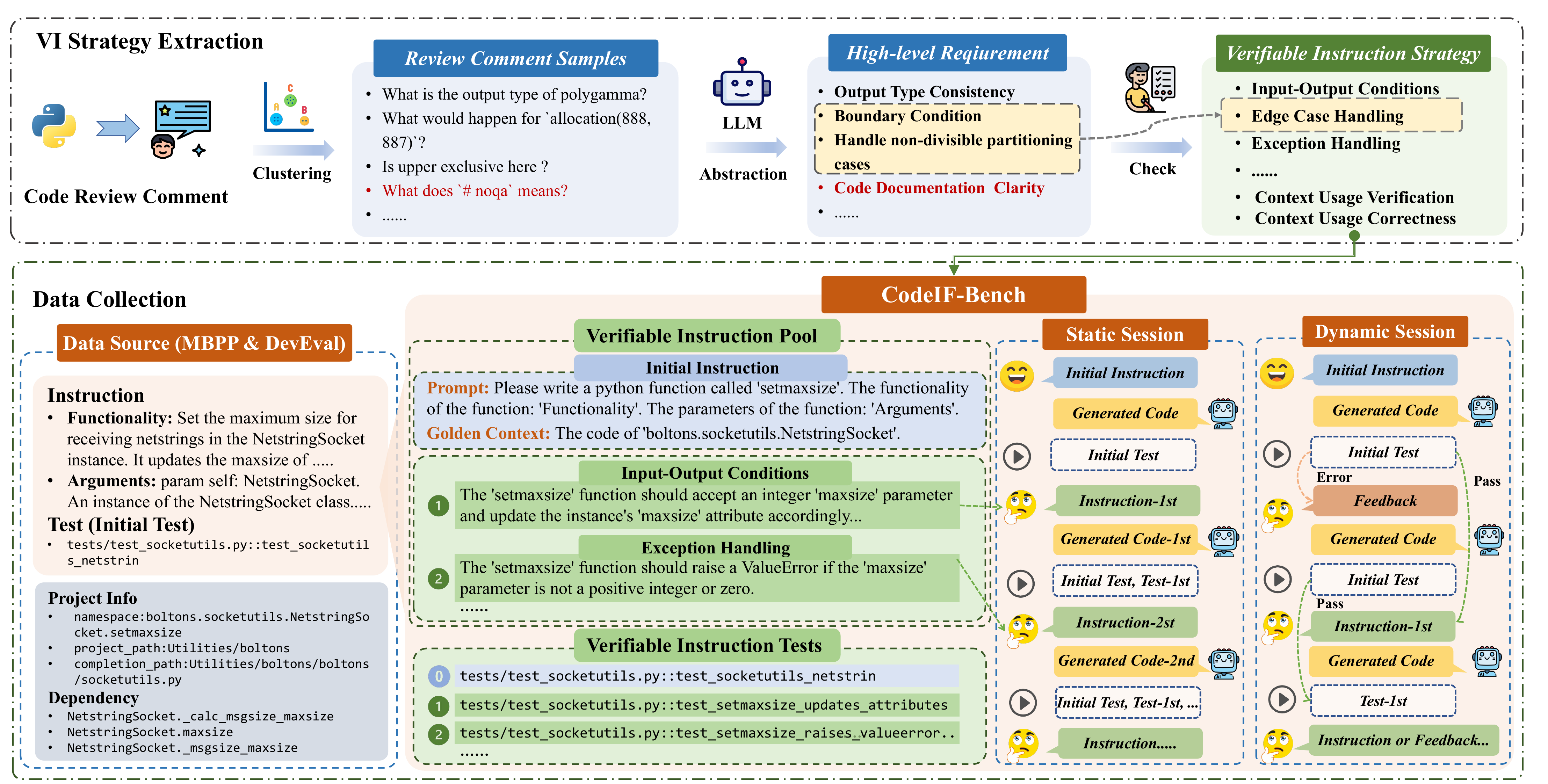}
	\caption{\bench construction pipeline. The top part illustrates the verifiable instruction strategy extraction process, and the bottom part presents the data collection procedure}
	\label{fig:total_process}
    \vspace{-0.5cm}
\end{figure*}

\subsection{Task Definition}
\label{task_def}
In interactive code generation, users may refine their requirements through multiple rounds of dialogue, and LLMs generate code to meet users' expectations and make adjustments based on feedback.
Specifically, for the $N$-th round, an LLM with strong instruction-following abilities can generate code that satisfies current instruction and the accumulated historical context from previous rounds.

For \bench, every data contains multiple \textbf{verifiable instructions (VI)} accompanied by test cases. Each verifiable instruction can be independently validated through tests to determine whether the model's output follows the instruction. By linking instructions to their corresponding tests, \bench evaluates the LLMs’ instruction-following capabilities throughout the interaction process by executing these tests. 
Specially, each data in \bench is denoted as $D_i = \{(I_1, T_1), (I_2, T_2), ...(I_k, T_k)\}$, $i\in\{1,2,...,|D|\},k\in\{1,2,...,|I|\},$ where $|D|$ is the total number of data and $|I|$ is the number of the verifiable instructions with tests. $I_k$ represents the $k$-th verifiable instruction, and $T_k$ represents the corresponding tests, which are both provided by the original benchmark or manual annotation. 
Based on \bench, we define the following two classic interactive code generation scenarios:

\ding{182} \textit{\textbf{Static Conversation}:} In this scenario, we simulate a linear progression of requirements where instruction sequences are predetermined. The LLM generates code for each instruction in sequence without iterative refinement, testing its ability to maintain consistency in multi-turn interactions without external feedback.
In the \textit{N-Round} conversation, we initialize the process at $N=1$ with the original programming instruction ($I_1$) from the benchmark. Subsequent instructions are then added sequentially from a predefined non-conflicting instruction sequence as $N$ increases.
In particular, for non-standalone tasks, which contain context-aware dependencies in repositories, $I_1$ consists of both the original task instruction and the golden contexts. Specifically, in $i$-th round, the input is formed as $\textit{Input} = (I_1,A_1,I_2,A_2,...,I_i)$, LLM's answer in this round is $\textit{A}_i$ evaluated with the test cases $(T_1,T_2,...,T_i)$.

\ding{183} \textit{\textbf{Dynamic Conversation}:} 
In real-world coding scenario, programmers typically review LLM-generated outputs to provide feedback. To simulate this human review process, we execute tests to obtain feedback and identify unmet requirements, thereby evaluating the LLM's IF capability in dynamic conversations. 
Specifically, each interaction round consists of both an instruction and its corresponding feedback, forming an instruction–feedback pair. We first initialize the instruction list containing all verifiable instructions and tests $(I_1,T_1, \dots ,I_n,T_n)$. During $i$ round, the system first executes the generated answer $A_{i-1}$ according to the current instruction list with tests. If $A_{i-1}$ fails to meet the requirements $(I_m,T_m)$, the instruction in this round is the $I_m$ with feedback $F_m$  to get the answer $A_i$. If $A_i$ passes the tests $T_m$, we keep the $(I_m,T_m)$ in current instruction list. Otherwise, we remove the $(I_m,T_m)$ from the instruction list. Until the instruction list is empty, the maximum number of interaction rounds is reached or the answer passes all instructions.

\begin{table*}[t]
\centering
\small
\setlength{\abovecaptionskip}{0.1cm}
\caption{Overview of the verifiable instruction strategies of \bench. Two context-related strategies marked with * are customized for repository-level coding task}
\label{tab:vi_type}
\resizebox{\textwidth}{!}{
\begin{tabular}{llccc}
\toprule
\textbf{Instruction Strategy} & \textbf{Description} & \textbf{L-1} & \textbf{L-2} & \textbf{L-3} \\
\midrule
Input-Output Conditions               & Validity checking of input/output parameters that the code expects to receive
& \ding{51} & \ding{51} & \ding{51}  \\
Exception Handling                   & Mechanisms for catching and handling exception (error) conditions 
& \ding{51} & \ding{51} & \ding{51}  \\
Edge Case Handling                   & Extreme boundary conditions for processing input data or operating scenarios           
& \ding{51} & \ding{51} & \ding{51}  \\
Functionality Extension              & Enhancement or extension of existing program functions               
& \ding{51} & \ding{51} & \ding{51}  \\
Annotation Coverage                  & The extent to which annotations are applied in the code
& \ding{51} & \ding{51} & \ding{51}  \\
Code Complexity                      & Control the complexity of the code structure (e.g., Cyclomatic Complexity \citep{Cyclomatic})
& \ding{51} & \ding{51} & \ding{51}  \\
Code Standard                        & Rules and guidelines for coding and structure (e.g., PEP 8 \citep{pep8})                       & \ding{51} & \ding{51} & \ding{51}  \\
Context Usage Verification*           & Check and ensure that contextual information is used in the code                         & \ding{55} & \ding{51} & \ding{51}  \\
Context Usage Correctness* & Check and ensure that contextual information is correctly used in the code
& \ding{55} & \ding{51} & \ding{51}  \\
\bottomrule
\end{tabular}
}
\vspace{-0.3cm}
\end{table*}

\subsection{Evaluation Metrics}\label{metrics}
In interactive code generation, we evaluate LLMs from three key perspectives. First, LLMs should not only follow the current instruction but also maintain consistent comprehension and execution across the entire dialogue. Second, we analyze the instruction forgetting phenomenon, where previously followed instructions are no longer respected in later turns. Third, in the \textit{Dynamic Conversation} setting, where feedback is introduced, we assess instruction-following efficiency—i.e., how many interaction rounds are required to complete $N$ instructions. Ideally, this would take exactly $N$ rounds, but due to corrective feedback, it may require $2N$ or more.

Based on these perspectives, we propose and report the following evaluation metrics: 
\begin{itemize}[leftmargin=*]
    \item \textbf{Instruction Accuracy (IA)} \citep{multiif}: This metric measures the percentage of instructions that the LLM accurately follows in \textbf{each round} of conversation. Specifically, in the $n$-th round of dialogue, given the historical dialogue and the current instruction $I_n$, the LLM outputs $A_n$. Whether the LLM complies with instruction $I_n$ is determined by judging whether $A_n$ passes the test $T_n$. 
    \begin{equation}
        IA = 
        \begin{cases}
            1, & \text{if } A_n \text{ passes the test } T_n \\
            0, & \text{if } A_n \text{ does not pass the test } T_n
        \end{cases}
    \end{equation}
    \vspace{-5pt}
    \item \textbf{Conversation Accuracy (CA)}: This metric measures the percentage of instructions followed by LLMs from the first turn to the current turn in current turn.
    Specifically, in the $n$-th round of dialogue, given the historical dialogue $\{I_1, A_1,...I_{n-1},A_{n-1}\}$ and the current instruction $I_n$, the LLM outputs $A_n$. We calculate the CA score in this turn by executing the test sequence $TS = \{T_1, ..., T_n\}$.
    \begin{equation}
        CA = \frac{\text{The Number of Tests Passed by }A_n \text{ in } TS}{\text{The Number of Tests in }TS}
    \end{equation}

    \item \textbf{Instruction Forgetting Ratio (IFR)} \citep{multiif}: This metric assesses the percentage of instructions that are followed previously and not followed subsequently.
    Specifically, an instruction is considered forgotten if it was followed in one of the previous turns ($1, 2, \dots, n-1$) but is not followed in the current turn ($n$). The test sequence passed in the previous turn $PTS = \{T'_1, ..., T'_k\}$ and the LLM
    outputs $A_n$.
    \begin{equation}
        IFR = \frac{\text{The Number of Tests Failed by }A_n \text{ in } PTS}{\text{The Number of Tests in }PTS}
    \end{equation}
    \item \textbf{Cumulative Instruction Following (CIF)}: This metric evaluates the cumulative instruction following of LLM in \textit{Dynamic Conversation}. It represents the number of instruction followed at the conclusion of the entire dynamic session. Specifically, in the last round of dialogue, given the test sequence $TS = \{T_1, ..., T_n\}$ in the data, the LLM outputs $A_{last}$. We calculate the CIF score in this turn by executing the test sequence.
    \begin{equation}
        CIF = \frac{\text{The Number of Tests Passed by }A_{last} \text{ in } TS}{\text{The Number of Tests in }TS}
    \end{equation}

\end{itemize}


\subsection{Benchmark Construction}\label{construction_pipeline}
In this section, we present the benchmark construction pipeline. To align the data in \bench with the distribution of real-world human instructions, as illustrated in Figure \ref{fig:total_process}, our dataset construction process consists of two key steps: \textbf{VI Strategy Extraction} and \textbf{Data Collection}. 
In \textbf{VI Strategy Extraction}, we first extract verifiable instruction strategies from actual code review comments, identifying common patterns and deriving high-level rules aligning the distribution of real-world to guide the LLM in producing verifiable instructions be independently verified by tests.
In \textbf{Data Collection}, we use VI Strategies to guide LLMs generating verifiable instructions based on initial programming tasks, and the generated instructions undergo rigorous manual evaluation and quality assurance procedures to ensure their reliability and validity.


\subsubsection{VI Strategy Extraction}
\label{Verifiable Instruction Strategy Extraction}
During software development, human code reviewers systematically assess code quality across multiple dimensions, subsequently providing actionable improvement suggestions. Developers then iteratively refine their code based on the review comment, forming a collaborative process known as code review \citep{codereview}. The code review process exhibits similarities to the interactive process between programmers and LLM-generated code. Inspired by this analogy, we leverage real-world code review comments \citep{codereview} to extract instruction strategies, identifying common patterns for introducing follow-up requirements, aligning the distribution of human instructions in the real world. However, not all instruction types can be independently verified through test cases. As demonstrated in Figure \ref{fig:total_process}, relatively subjective instructions (\textit{e.g.}, ``What does \# noqa mean?'') lack objective evaluation criteria, making them difficult to assess via automated testing. Furthermore, determining whether an instruction type can produce objective, verifiable outputs is non-trivial and often requires deep domain expertise. To address this, we employ manual annotation to filter verifiable instruction types serving as guidelines for generating diverse and actionable, verifiable instructions.


To extract diverse verifiable strategies, we first employ the KCenterGreedy \citep{kcenter} algorithm, which has demonstrated effectiveness in selecting a well-distributed set of representative samples based on text embeddings, to select 500 candidate review comment examples. (In our experiment, 500 is far greater than the actual number of unique instructions.) Then, we prompt GPT-4o to generate strategies for each candidate comment, which we refer to as ``High-level Requirement'' to guide LLM to generate verifiable instructions. 
To guarantee the validity and verifiability of these strategies, we invited 2 experienced software engineering experts with at least 4 years of programming experience to check these strategies, by removing unverifiable instruction types, merging similar ones, and refining ambiguous requirements into high-quality VI strategies. All ambiguous samples were resolved through discussion and mutual agreement to ensure strict adherence to the above criteria.

Finally, we obtain 9 verifiable instruction strategies, covering functional requirements (such as \textit{Exception Handling}, \textit{Input-Output Conditions}, \textit{etc}) and non-functional requirements (such as \textit{Annotation Coverage}, \textit{Code Standard}, \textit{etc}). Among these strategies, two are context-related: \textit{i.e.}, \textit{Context Usage Verification} and \textit{Context Usage Correctness}, supporting the assessment of whether LLMs can effectively and accurately leverage the provided project-specific contexts for repository-level code generation tasks. Detailed information on VI strategies is illustrated in Table \ref{tab:vi_type}.

\begin{table*}[t]
\centering
\small
\setlength{\abovecaptionskip}{0.1cm}
\caption{The statistics of \bench. SA stands for StandAlone. Non-SA stands for repository-level code generation. Avg. L is the average lengths (tokens) of instructions calculated by GPT's tokenizer 
}
\label{tab:benchmark_overview}
\resizebox{\textwidth}{!}{
\begin{tabular}{l|c|c|cc|cc|ccc|cc}
\toprule
\multirow{2}{*}{\textbf{Level}} & \multirow{2}{*}{\textbf{\#Task}} & \multirow{2}{*}{\textbf{\#Repo}} & \multicolumn{2}{c|}{\textbf{Code Type}} & \multicolumn{2}{c|}{\textbf{Dependency}} & \multicolumn{3}{c|}{\textbf{Verifiable Instruction}} & \multicolumn{2}{c}{\textbf{Avg. L}} \\
 & ~ & & ~ SA (\%) & Non-SA (\%) & Type & \#Total  & \#Total & \#Type & Test  
 & Base  & VI  \\ 
\midrule
L-1 & 70 & 14 & 100\% & 0\% & SA & 0  & 487 & 7 &\ding{51} & 43.8 & 17.2\\
L-2 & 40 & 26 &  0\% & 100\% & intra\_file & 126 & 360 & 9 &\ding{51} & 9.6K & 30.0\\
L-3 & 14 & 11 & 0\% & 100\% & cross\_file & 98 & 126 & 9 &\ding{51} & 28K & 30.0\\

\midrule
Total  & 124 & 42 & 56.5\% & 43.5\% & All & 224 & 964 & 9 &\ding{51} & 7K & 22.3 \\

\bottomrule
\end{tabular}
}
\vspace{-0.3cm}
\end{table*}
\subsubsection{Data Collection}
\label{Benchmark Construction}

To construct \bench, we first sample 124 programming tasks from two widely-used code generation benchmarks, where 50 tasks from MBPP \citep{MBPP} and 74 tasks from DevEval \citep{DevEval}, encompassing both standalone functions and repository-level code generation. Specifically, for standalone function, we randomly sample 50 programming tasks 
from MBPP test dataset. The repo-level code generation tasks are sourced from DevEval, where the programming tasks are collected from 115 real-world projects across 10 domains covering different dependency types:
\begin{itemize}[leftmargin=*]
    \item \textbf{Level-1 (L-1)}: Programming tasks that rely solely on built-in functions and standard libraries.
    \item \textbf{Level-2 (L-2)}: Programming tasks that require intra-file or intra-class context.
    \item\textbf{Level-3 (L-3)}: Programming tasks that require both intra-file and cross-file context.
\end{itemize}

We select data based on the 10 repository classes and dependency types provided by DevEval, randomly selecting 2 programming tasks from each class, resulting in 20 programming tasks for each dependency type. Due to the limited number of cross-file programming tasks, resulting in only 14 data points. Ultimately, we obtain 74 repository-level programming tasks.
Then we generate the \textit{VIs} for each task based on the extracted instruction strategies, and we also build corresponding unit tests ($T$) for assessing whether the code follows the \textit{VI}.
Inspired by \citep{ou2024inductive}, the instructions and tests are produced by LLMs, followed by human review and verification through execution.
Specifically, the extracted \textit{VI} strategies are applied to each task to guide the generation of verifiable instructions (7 for SA and 9 for Non-SA). Then we prompt GPT-4o to generate ($\textit{VI}$, $T$) pairs based on the initial instruction and VI strategies. 
To further ensure data quality, we engaged two developers with over 4 years of Python programming experience to individually check and refine the data. They refined the dataset based on the following two criteria: 1. Ensuring \textit{instruction description accuracy}—identifying vague descriptions or logical conflicts with the original programming task; 2. Validating \textit{test correctness}—ensuring the test can verify instructions and introducing necessary dependencies to ensure proper execution of test functions. Through meticulous examination, they identified over 55\% of the samples requiring refinement and achieved a Cohen's Kappa score of 0.73. All ambiguous samples were resolved through discussion and mutual agreement to ensure strict adherence to the above criteria.

\subsection{Dataset Overview}
\begin{table*}[t]
\centering
\small
\setlength{\abovecaptionskip}{0.1cm}
\caption{The comparison between popular benchmarks and \bench}
\label{tab:benchmark_comparison}
\resizebox{\textwidth}{!}{
\begin{tabular}{lcccc}
\toprule
\textbf{Benchmark} & \textbf{Code Domain} & \textbf{Real World Scenario} & \textbf{Robust Metric} & \textbf{Multi-Round} \\
\midrule
\rowcolor[rgb]{.741,.843,.933}
\multicolumn{5}{c}{\textbf{Code Generation Benchmarks}} \\
\midrule
HumanEval~\citep{HumanEval}   & \ding{51} & \ding{55} & \ding{51} & \ding{55} \\
MBPP~\citep{MBPP}             & \ding{51} & \ding{55} & \ding{51} & \ding{55} \\
CoderEval~\citep{CoderEval}   & \ding{51} & \ding{51} & \ding{51} & \ding{55} \\
DevEval~\citep{DevEval}       & \ding{51} & \ding{51} & \ding{51} & \ding{55} \\
RepoEval~\citep{RepoEval}     & \ding{51} & \ding{51} & \ding{55} & \ding{55} \\
CanItEdit~\citep{CanItEdit}   & \ding{51} & \ding{55} & \ding{51} & \ding{55} \\
\midrule
\rowcolor[rgb]{.741,.843,.933}
\multicolumn{5}{c}{\textbf{NLP Instruction-Following Benchmarks}} \\
\midrule
MT-Bench~\citep{MT-Bench}     & \ding{55} & \ding{55} & \ding{55} & \ding{51} \\
Multi-IF~\citep{multiif}      & \ding{55} & \ding{55} & \ding{51} & \ding{51} \\
InFo-Bench~\citep{InfoBench}  & \ding{55} & \ding{55} & \ding{55} & \ding{55} \\
\midrule
CodeIF~\citep{codeif}         & \ding{51} & \ding{55} & \ding{51} & \ding{55} \\
\textbf{\bench}              & \ding{51} & \ding{51} & \ding{51} & \ding{51} \\
\bottomrule
\end{tabular}
}
\vspace{-0.3cm}
\end{table*}

As shown in Table \ref{tab:benchmark_overview}, we finally construct \bench, which consists of 124 programming tasks along with 964 verifiable instructions. 
The tasks in \bench can be categorized into 3 levels based on the scope of context usage (dependency). 
Table \ref{tab:benchmark_comparison} shows a comparison between \bench and existing benchmarks:

\noindent \textbf{Code Domain \& Real World Scenario.} \bench supports for both standalone function-level and repository-level code generation evaluation, where the programming tasks are derived from real-world projects \citep{DevEval} and crowd-sourced problems \citep{MBPP}.


\noindent \textbf{Robust Metric.} \bench enables an accurate evaluation of instruction following ability by calculating execution-based scores with corresponding test cases rather than LLM-as-Judge.

\noindent \textbf{Multi-Round Support.} \bench enables the evaluation of LLMs' ability to follow multi-turn instructions. Specifically, we constructed two types of dialogue scenarios—\textit{Static Conversation} and \textit{Dynamic Conversation} with feedback—to closely resemble real-world situations.

\section{Evaluation}
\subsection{Research Questions}
In this paper, we aim to answer the following research questions:
\begin{itemize}[leftmargin=*]
    \item \noindent\textbf{RQ1: Performance in Static Conversation}. How does LLM perform in instruction-following under \textit{Static Conversation}?
    \item \noindent\textbf{RQ2: Performance in  Dynamic Conversation}. How does LLM perform in instruction-following under \textit{Dynamic Conversation}?
    \item \noindent\textbf{RQ3: Performance on Various VI Strategies}. How does LLM perform in instruction-following on different strategies of instructions?
    \item \noindent\textbf{RQ4: Exploration of IF Enhancement}. How do prompt-enhancement methods perform in enhancing instruction-following capabilities in interactive coding tasks?
\end{itemize}
\subsection{Experimental Settings}
\label{Experiment Settings}

As mentioned in Section \ref{task_def}, \bench supports two scenarios of \textit{N-Round} dialogue evaluation. 
For \textbf{\textit{Static Conversation}}, the first round consists of the initial programming task, with verifiable instructions introduced in subsequent rounds. Since independent instructions could potentially conflict (\textit{e.g.}, ``Functionality Extension'' instructions might interfere with existing code functionality), we predefined a set of non-conflicting instruction sequences to mitigate this issue (\textit{e.g.}, the 'Functionality Extension' instruction is added at the end of the conversation). 
In this paper, we report the experimental results based on the predefined instruction sequence. 

For \textbf{\textit{Dynamic Conversation}}, the initial programming task serves as the instruction of the first round, and then we execute the tests in the instruction list (the initial state is all verifiable instructions) to select unfulfilled instructions with feedback in the next round. 
Next, we execute tests to verify whether the LLM-generated responses satisfy the current instruction. If the response fails, the current instruction is removed from the instruction list; otherwise, it is retained. This process continues until the instruction list becomes empty, the maximum number of interaction rounds is reached, or all instruction are successfully satisfied. In our experiments, the maximum number of interaction rounds is set to 9 for SA and 11 for Non-SA.

Due to limited budget and LLMs' poor performance in L-3 (our preliminary experiments found that the most LLMs failed on L-3 tasks), only L-1 and L-2 tasks are included in the experiment. To ensure deterministic outputs for the evaluation, we employ the greedy decoding strategy, \textit{i.e.}, we set the parameters to ``temperature=0''.

\subsection{Evaluated LLMs}
We select the following SOTA LLMs for code generation tasks: 
\begin{itemize}[leftmargin=*]
    \item \textbf{Closed-Source LLM}: We chose Claude-3.5-Sonnet \citep{claude} and GPT-4o \citep{gpt4o}, powerful coding models in the field of code generation.
    \item \textbf{Open-Source LLM}: We select one of the open-source large-scale SOTA LLMs Deepseek-V3 \citep{zhu2024deepseek}, and open-source small-scale SOTA code LLMs Qwen2.5-Coder-Instruct-7, 14, 32B \citep{qwencoder}.
\end{itemize}
\section{Results and Analysis}

\subsection{RQ1: Performance in \textit{Static Conversation}}

\begin{table*}[t]
\centering
\small
\setlength{\abovecaptionskip}{0.1cm}
\caption{IA results in Static Conversation. Values with darker colors indicate higher performance}
\label{tab:IA_RQ1}
\resizebox{\textwidth}{!}{
\begin{tabular}{l|l|ccccccccccc}
\toprule
\textbf{Level} & \textbf{Model} & Turn-1 & Turn-2 & Turn-3 & Turn-4 & Turn-5 & Turn-6 & Turn-7 & Turn-8 & Turn-9 & Turn-10 & Avg. \\
\midrule
 \multirow{6}{*}{L1} & Claude-3.5-Sonnet & \scorecellgreen{58.6} & \scorecellgreen{84.3} & \scorecellgreen{84.3} & \scorecellgreen{75.7} & \scorecellgreen{70.0} & \scorecellgreen{82.1} & \scorecellgreen{52.0} & \scorecellgreen{49.0} & - & - & \scorecellgreen{69.5} \\
& GPT-4o & \scorecellgreen{57.1} & \scorecellgreen{80.7} & \scorecellgreen{78.5} & \scorecellgreen{76.7} & \scorecellgreen{74.9} & \scorecellgreen{70.5} & \scorecellgreen{58.7} & \scorecellgreen{46.9} & - & - & \scorecellgreen{64.0} \\
& DeepSeek-V3 & \scorecellgreen{60.0} & \scorecellgreen{80.0} & \scorecellgreen{84.3} & \scorecellgreen{77.4} & \scorecellgreen{55.7} & \scorecellgreen{71.8} & \scorecellgreen{29.1} & \scorecellgreen{54.9} & - & - & \scorecellgreen{64.2} \\
& Qwen2.5-Coder-7B & \scorecellgreen{44.3} & \scorecellgreen{65.7} & \scorecellgreen{62.9} & \scorecellgreen{58.6} & \scorecellgreen{35.7} & \scorecellgreen{66.0} & \scorecellgreen{11.7} & \scorecellgreen{41.5} & - & - & \scorecellgreen{48.3} \\
& Qwen2.5-Coder-14B & \scorecellgreen{48.6} & \scorecellgreen{72.9} & \scorecellgreen{70.0} & \scorecellgreen{64.3} & \scorecellgreen{40.0} & \scorecellgreen{72.0} & \scorecellgreen{17.4} & \scorecellgreen{44.5} & - & - & \scorecellgreen{53.7} \\
& Qwen2.5-Coder-32B & \scorecellgreen{51.4} & \scorecellgreen{75.7} & \scorecellgreen{ 75.7} & \scorecellgreen{68.6} & \scorecellgreen{50.0} & \scorecellgreen{69.0} & \scorecellgreen{17.5} & \scorecellgreen{41.7} & - & - & \scorecellgreen{56.2} \\
\midrule
 \multirow{6}{*}{L2} & Claude-3.5-Sonnet & \scorecellgreen{55.0} & \scorecellgreen{60.0} & \scorecellgreen{62.5} & \scorecellgreen{52.5} & \scorecellgreen{37.5} & \scorecellgreen{25.0} & \scorecellgreen{7.5} & \scorecellgreen{50.0} & \scorecellgreen{30.0} & \scorecellgreen{37.5} & \scorecellgreen{41.8} \\
& GPT-4o & \scorecellgreen{47.5} & \scorecellgreen{50.0} & \scorecellgreen{50.0} & \scorecellgreen{42.5} & \scorecellgreen{32.5} & \scorecellgreen{37.5} & \scorecellgreen{2.5} & \scorecellgreen{37.5} & \scorecellgreen{30.0} & \scorecellgreen{20.0} & \scorecellgreen{35.0} \\
& DeepSeek-V3 & \scorecellgreen{50.0} & \scorecellgreen{55.0} & \scorecellgreen{62.5} & \scorecellgreen{52.5} & \scorecellgreen{27.5} & \scorecellgreen{30.0} & \scorecellgreen{2.5} & \scorecellgreen{45.5} & \scorecellgreen{45.5} & \scorecellgreen{27.5} & \scorecellgreen{39.8} \\
& Qwen2.5-Coder-7B & \scorecellgreen{27.5} & \scorecellgreen{32.5} & \scorecellgreen{42.5} & \scorecellgreen{22.5} & \scorecellgreen{22.5} & \scorecellgreen{20.0} & \scorecellgreen{2.5} & \scorecellgreen{20.0} & \scorecellgreen{12.5} & \scorecellgreen{12.5} & \scorecellgreen{21.5} \\
& Qwen2.5-Coder-14B & \scorecellgreen{47.5} & \scorecellgreen{47.5} & \scorecellgreen{47.5} & \scorecellgreen{42.5} & \scorecellgreen{25.0} & \scorecellgreen{32.5} & \scorecellgreen{2.5} & \scorecellgreen{30.0} & \scorecellgreen{27.5} & \scorecellgreen{25.0} & \scorecellgreen{32.8} \\
& Qwen2.5-Coder-32B & \scorecellgreen{42.5} & \scorecellgreen{37.5} & \scorecellgreen{35.0} & \scorecellgreen{27.5} & \scorecellgreen{12.5} & \scorecellgreen{15.0} & \scorecellgreen{2.5} & \scorecellgreen{15.0} & \scorecellgreen{7.5} & \scorecellgreen{2.5} & \scorecellgreen{19.8} \\
\bottomrule
\end{tabular}
}
\vspace{-0.3cm}
\end{table*}

\begin{table*}[t]
\centering
\small
\setlength{\abovecaptionskip}{0.1cm}
\caption{CA results in Static Conversation. Values with darker colors indicate higher performance}
\label{tab:CA_RQ1}
\resizebox{\textwidth}{!}{
\begin{tabular}{l|l|cccccccccc}
\toprule
\textbf{Level} & \textbf{Model} & Turn-1 & Turn-2 & Turn-3 & Turn-4 & Turn-5 & Turn-6 & Turn-7 & Turn-8 & Turn-9 & Avg. \\
\midrule
 \multirow{6}{*}{L1} & Claude-3.5-Sonnet & \scorecellblue{58.6} & \scorecellblue{79.8} & \scorecellblue{80.3} & \scorecellblue{80.2} & \scorecellblue{78.3} & \scorecellblue{76.2} & \scorecellblue{68.0} & - & - & \scorecellblue{74.5} \\
& GPT-4o & \scorecellblue{57.1} & \scorecellblue{80.7} & \scorecellblue{78.5} & \scorecellblue{76.7} & \scorecellblue{74.9} & \scorecellblue{70.5} & \scorecellblue{ 58.7} & - & - & \scorecellblue{71.0} \\
& DeepSeek-V3 & \scorecellblue{60.0} & \scorecellblue{77.6} & \scorecellblue{78.0} & \scorecellblue{77.4} & \scorecellblue{72.8} & \scorecellblue{62.4} & \scorecellblue{57.4} & - & - & \scorecellblue{69.4} \\
& Qwen2.5-Coder-7B & \scorecellblue{44.3} & \scorecellblue{59.8} & \scorecellblue{63.1} & \scorecellblue{60.0} & \scorecellblue{56.2} & \scorecellblue{54.1} & \scorecellblue{46.7} & - & - & \scorecellblue{54.9} \\
& Qwen2.5-Coder-14B & \scorecellblue{48.6} & \scorecellblue{70.0} & \scorecellblue{68.8} & \scorecellblue{65.6} & \scorecellblue{62.7} & \scorecellblue{60.8} & \scorecellblue{55.8} & - & - & \scorecellblue{61.7} \\
& Qwen2.5-Coder-32B & \scorecellblue{51.4} & \scorecellblue{74.1} & \scorecellblue{74.5} & \scorecellblue{72.5} & \scorecellblue{68.5} & \scorecellblue{61.8} & \scorecellblue{57.0} & - & - & \scorecellblue{65.7} \\
\midrule
 \multirow{6}{*}{L2} & Claude-3.5-Sonnet & \scorecellblue{55.0} & \scorecellblue{55.0} & \scorecellblue{55.8} & \scorecellblue{55.6} & \scorecellblue{52.0} & \scorecellblue{44.6} & \scorecellblue{39.3} & \scorecellblue{34.1} & \scorecellblue{25.8} & \scorecellblue{46.4} \\
& GPT-4o & \scorecellblue{47.5} & \scorecellblue{51.2} & \scorecellblue{48.3} & \scorecellblue{46.9} & \scorecellblue{35.0} & \scorecellblue{36.3} & \scorecellblue{31.1} & \scorecellblue{29.7} & \scorecellblue{27.8} & \scorecellblue{39.3} \\
& DeepSeek-V3 & \scorecellblue{50.0} & \scorecellblue{51.2} & \scorecellblue{54.2} & \scorecellblue{51.2} & \scorecellblue{35.0} & \scorecellblue{32.1} & \scorecellblue{29.3} & \scorecellblue{31.6} & \scorecellblue{33.6} & \scorecellblue{40.9} \\
& Qwen2.5-Coder-7B & \scorecellblue{27.5} & \scorecellblue{26.3} & \scorecellblue{30.8} & \scorecellblue{21.3} & \scorecellblue{20.0} & \scorecellblue{16.3} & \scorecellblue{14.3} & \scorecellblue{15.0} & \scorecellblue{13.1} & \scorecellblue{20.5} \\
& Qwen2.5-Coder-14B & \scorecellblue{47.5} & \scorecellblue{45.0} & \scorecellblue{44.2} & \scorecellblue{43.1} & \scorecellblue{30.0} & \scorecellblue{32.5} & \scorecellblue{27.9} & \scorecellblue{26.3} & \scorecellblue{25.8} & \scorecellblue{35.8} \\
& Qwen2.5-Coder-32B & \scorecellblue{42.5} & \scorecellblue{35.0} & \scorecellblue{35.0} & \scorecellblue{29.4} & \scorecellblue{19.0} & \scorecellblue{15.0} & \scorecellblue{11.1} & \scorecellblue{8.4} & \scorecellblue{6.1} & \scorecellblue{22.4} \\
\bottomrule
\end{tabular}
}
\vspace{-0.3cm}
\end{table*}

To answer this research question, we evaluated the LLMs on \bench under the \textit{Static Conversation} setting. 
The Instruction Accuracy (IA) results are summarized in Table~\ref{tab:IA_RQ1}. In the L-1 task, most LLMs demonstrated strong abilities to follow users’ immediate instructions, with IA averages exceeding 50\%. Notably, Qwen2.5-Coder-32B-Instruct achieved performance comparable to the large open-source model DeepSeek-V3. However, in the L-2 task, substantial performance disparities emerged across models. For example, Claude successfully completed 60\% of the tasks in the first round, whereas Qwen2.5-Coder-7B-Instruct—one of the strongest open-source models—achieved a completion rate below 30\%. Interestingly, Qwen2.5-Coder-14B-Instruct achieved first-round performance comparable to both the closed-source GPT-4o and the open-source SOTA model DeepSeek-V3. This finding suggests that carefully optimized training data and algorithms can enable smaller-parameter models to match larger open-source counterparts in single-round tasks. With increasing dialogue rounds, all models show reduced instruction-following capability, with the Qwen series experiencing a notably sharper decline than stronger models like Claude. This highlights the performance gap between open-source and closed-source models in interactions. These L-1 and L-2 results indicate that long-context management critically affects instruction following, and more capable LLMs better resist such long-context-induced degradation. It is worth noting that the performance differences among LLMs were most pronounced in Rounds 5, 7, and 8, likely due to the instruction types. This phenomenon is further examined in section \ref{RQ3}.

Notably, Qwen2.5-Coder-14B-Instruct outperformed the 32B variant and was only marginally weaker than DeepSeek-V3 and GPT-4o. We observed that the 32B model often produced format errors during later dialogue turns—for instance, generating patch-style outputs while neglecting the initial instruction’s function requirement. These observations suggest that sustained instruction-following performance in multi-turn dialogues is primarily determined by a model’s exposure to multi-turn supervision rather than by parameter scale alone. Although Qwen2.5-Coder-32B-Instruct possesses greater capacity, its weaker performance relative to the 14B variant indicates that insufficient optimization for multi-turn instruction consistency can undermine its potential. This underscores that training strategies emphasizing dialogue-level instruction alignment are essential for maintaining stable adherence across turns. 

\begin{figure}[t]
    \centering

    \begin{subfigure}[b]{0.48\linewidth}
        \includegraphics[width=\linewidth]{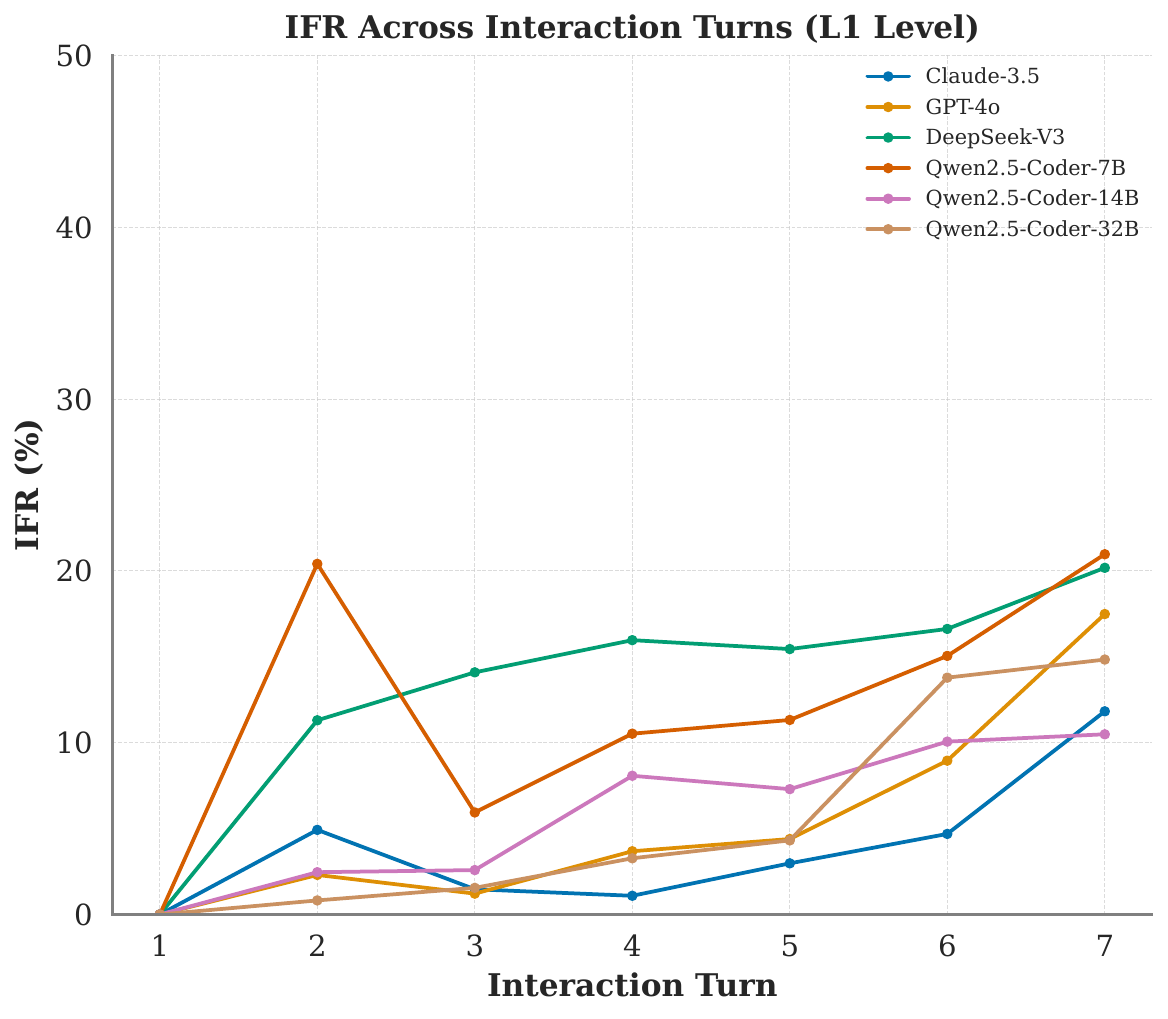}
        \caption{L1-Static Conversation}
        \label{fig:L1}
    \end{subfigure}
    \hfill
    \begin{subfigure}[b]{0.48\linewidth}
        \includegraphics[width=\linewidth]{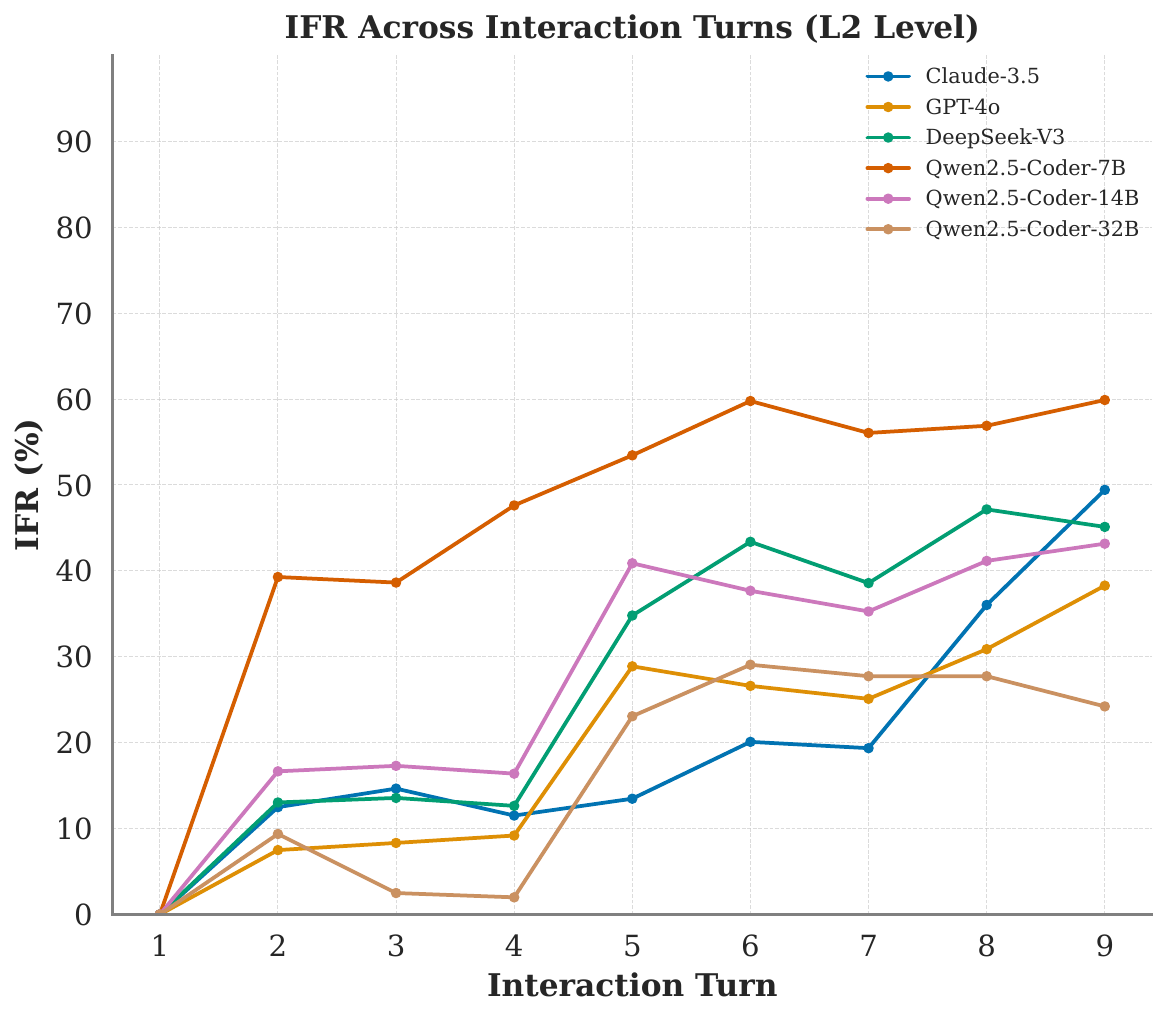}
        \caption{L2-Static Conversation}
        \label{fig:dev_turn}
    \end{subfigure}

    \caption{IFR results in Static Conversation}
    \label{fig:RQ1_IFR}
\end{figure}
In the L-1 task, the Conversation Accuracy (CA) of all LLMs exhibits an initial rise followed by a gradual decline as dialogue rounds progress. This pattern suggests that LLMs can effectively follow both the user’s current instruction and prior instructions during the early interaction stages. The subsequent degradation can be attributed to two main factors.
First, as the number of instructions and the volume of conversational history increase, models experience a reduced ability to follow the current instruction, as reflected by declining IA scores.
Second, Figure~\ref{fig:RQ1_IFR} presents the instruction forgetting rate (IFR) results, showing that most models display a progressive increase in IFR with longer dialogues—indicating that LLMs tend to “forget” earlier instructions. This manifests in erroneous edits to previously completed instructions and poor control over the editing scope.

In the L-2 repository task, the introduction of additional repository context (averaging 9.8K tokens) led to a pronounced drop in CA metrics for nearly all models as the dialogue advanced. This was accompanied by further decreases in IA and substantial increases in IFR. These results reinforce that the repository context is a critical determinant of LLM performance during multi-turn interactions, exposing persistent limitations in instruction-following robustness under real-world development scenarios.

\begin{RQbox} \textbf{RQ1 Summary: }
Under the Static Conversation setting, both Instruction Accuracy (IA) and Conversation Accuracy (CA) decline as dialogue rounds increase: in L-1, CA rises initially but then drops, while IA decreases with accumulating instructions. Meanwhile, Instruction Forgetting Rate (IFR) steadily grows, indicating progressive forgetting and erroneous edits in multi-turn interactions. Adding repository context (L-2, about 9.8K tokens) amplifies these effects—CA and IA fall further, and IFR increases sharply. 

\end{RQbox}

\subsection{RQ2: Performance in \textit{Dynamic Conversation}}

\begin{table*}[t]
\centering
\small
\setlength{\abovecaptionskip}{0.1cm}
\caption{IA results in Dynamic Conversation. Values with darker colors indicate higher performance}
\label{tab:RQ2_IA}
\resizebox{\textwidth}{!}{
\begin{tabular}{l|l|cccccccccccc}
\toprule
\textbf{Level} & \textbf{Model} & Turn 1 & Turn 2 & Turn 3 & Turn 4  & Turn 5 & Turn 6  & Turn 7 & Turn 8 & Turn 9 & Turn 10 & Turn 11 & Avg. \\
\midrule
 \multirow{6}{*}{L1} & Claude-3.5-Sonnet & \scorecellgreen{55.7} & \scorecellgreen{71.4} & \scorecellgreen{78.8} & \scorecellgreen{66.1} & \scorecellgreen{60.0} & \scorecellgreen{57.1} & \scorecellgreen{56.5} & \scorecellgreen{50.0} & \scorecellgreen{66.7} & - & - & \scorecellgreen{68.2} \\
& GPT-4o & \scorecellgreen{57.1} & \scorecellgreen{54.3} & \scorecellgreen{64.2} & \scorecellgreen{55.7} & \scorecellgreen{58.8} & \scorecellgreen{58.1} & \scorecellgreen{42.9} & \scorecellgreen{60.0} & \scorecellgreen{50.0}   & - & - & \scorecellgreen{59.0} \\
& DeepSeek-V3 & \scorecellgreen{55.7} & \scorecellgreen{60.0} & \scorecellgreen{56.5} & \scorecellgreen{64.4} & \scorecellgreen{59.2} & \scorecellgreen{53.3} & \scorecellgreen{65.2} & \scorecellgreen{53.8}  & \scorecellgreen{63.6}  & - & & \scorecellgreen{60.2}\\
& Qwen2.5-Coder-7B & \scorecellgreen{48.6} & \scorecellgreen{41.4} & \scorecellgreen{44.1} & \scorecellgreen{39.3} & \scorecellgreen{35.2} & \scorecellgreen{36.4} & \scorecellgreen{27.6} & \scorecellgreen{26.9} & \scorecellgreen{25.0} & - & -& \scorecellgreen{43.9} \\
& Qwen2.5-Coder-14B & \scorecellgreen{45.7} & \scorecellgreen{51.4} & \scorecellgreen{52.9} & \scorecellgreen{45.9} & \scorecellgreen{45.8} & \scorecellgreen{50.0} & \scorecellgreen{37.5}  & \scorecellgreen{44.4}& \scorecellgreen{50.0}  & - & -& \scorecellgreen{50.0}  \\
& Qwen2.5-Coder-32B & \scorecellgreen{55.7} & \scorecellgreen{52.9} & \scorecellgreen{51.5} & \scorecellgreen{54.1} & \scorecellgreen{47.1} & \scorecellgreen{39.4} & \scorecellgreen{35.0}  & \scorecellgreen{62.5}& \scorecellgreen{35.7}  & - & -  & \scorecellgreen{52.3} \\

\midrule
 \multirow{6}{*}{L2} & Claude-3.5-Sonnet & \scorecellgreen{52.5} & \scorecellgreen{45.0} & \scorecellgreen{42.5} & \scorecellgreen{40.0} & \scorecellgreen{28.9} & \scorecellgreen{14.7} & \scorecellgreen{29.0} & \scorecellgreen{25.0} & \scorecellgreen{26.3} & \scorecellgreen{33.3} & \scorecellgreen{27.3} & \scorecellgreen{37.2}\\
& GPT-4o & \scorecellgreen{47.5} & \scorecellgreen{32.5} & \scorecellgreen{40.0} & \scorecellgreen{25.0} & \scorecellgreen{28.9} & \scorecellgreen{21.6} & \scorecellgreen{42.9} & \scorecellgreen{35.7} & \scorecellgreen{27.3} & \scorecellgreen{23.8}  & \scorecellgreen{22.2} & \scorecellgreen{33.9} \\
& DeepSeek-V3 & \scorecellgreen{50.0} & \scorecellgreen{47.5} & \scorecellgreen{35.0} & \scorecellgreen{35.0} & \scorecellgreen{28.9}  & \scorecellgreen{43.8} & \scorecellgreen{33.3} & \scorecellgreen{54.5} & \scorecellgreen{42.1} & \scorecellgreen{58.8} & \scorecellgreen{38.5} & \scorecellgreen{41.8} \\
& Qwen2.5-Coder-7B & \scorecellgreen{35.0} & \scorecellgreen{10.0} & \scorecellgreen{20.0} & \scorecellgreen{20.0} & \scorecellgreen{10.0} & \scorecellgreen{17.9} & \scorecellgreen{11.1} & \scorecellgreen{14.3} & \scorecellgreen{20.0} & \scorecellgreen{18.8}  & \scorecellgreen{19.4} & \scorecellgreen{18.8}\\
& Qwen2.5-Coder-14B & \scorecellgreen{47.5} & \scorecellgreen{32.5} & \scorecellgreen{25.0} & \scorecellgreen{25.0} & \scorecellgreen{17.9} & \scorecellgreen{14.3} & \scorecellgreen{23.5} & \scorecellgreen{14.8} & \scorecellgreen{25.9} & \scorecellgreen{25.9} & \scorecellgreen{20.8}& \scorecellgreen{27.3}\\
& Qwen2.5-Coder-32B & \scorecellgreen{40.0} & \scorecellgreen{27.5} & \scorecellgreen{17.5} & \scorecellgreen{12.5} & \scorecellgreen{12.8} & \scorecellgreen{0.0} & \scorecellgreen{2.7} & \scorecellgreen{5.7} & \scorecellgreen{2.9} & \scorecellgreen{0.0} & \scorecellgreen{0.0}& \scorecellgreen{14.3}\\
\bottomrule
\end{tabular}
}
\vspace{-0.3cm}
\end{table*}

\begin{table*}[t]
\centering
\small
\setlength{\abovecaptionskip}{0.1cm}
\caption{CA and CIF results in Dynamic Conversation. Values with darker colors indicate higher performance}
\label{tab:RQ2_CA}
\resizebox{\textwidth}{!}{
\begin{tabular}{l|l|ccccccccccc}
\toprule
\textbf{Level} & \textbf{Model} & Turn 1 & Turn 2 & Turn 3 & Turn 4  & Turn 5 & Turn 6  & Turn 7 & Turn 8 & Turn 9 & Turn 10 & CIF \\
\midrule
 \multirow{6}{*}{L1} & Claude-3.5-Sonnet & \scorecellblue{48.6} & \scorecellblue{58.7} & \scorecellblue{66.5} & \scorecellblue{65.8} & \scorecellblue{64.6} & \scorecellblue{58.8} & \scorecellblue{50.4} & \scorecellblue{35.7}  & - & -& \scorecellblue{74.5} \\
& GPT-4o & \scorecellblue{49.0} & \scorecellblue{55.3} & \scorecellblue{60.7} & \scorecellblue{59.8} & \scorecellblue{61.3} & \scorecellblue{57.1} & \scorecellblue{35.5} & \scorecellblue{44.8} & - & - & \scorecellblue{71.9} \\
& DeepSeek-V3 & \scorecellblue{48.6} & \scorecellblue{58.0} & \scorecellblue{63.8} & \scorecellblue{62.3} & \scorecellblue{65.1} & \scorecellblue{55.2} & \scorecellblue{52.0} & \scorecellblue{30.8}  & - & -& \scorecellblue{72.3} \\
& Qwen2.5-Coder-7B & \scorecellblue{44.8} & \scorecellblue{48.0} & \scorecellblue{50.3} & \scorecellblue{50.0} & \scorecellblue{41.7} & \scorecellblue{35.4} & \scorecellblue{19.3} & \scorecellblue{22.8} & - & - & \scorecellblue{52.3} \\
& Qwen2.5-Coder-14B & \scorecellblue{46.1} & \scorecellblue{54.2} & \scorecellblue{58.3} & \scorecellblue{57.2} & \scorecellblue{54.2} & \scorecellblue{52.1} & \scorecellblue{38.8}  & \scorecellblue{32.3}  & - & -& \scorecellblue{65.0} \\
& Qwen2.5-Coder-32B & \scorecellblue{47.2} & \scorecellblue{56.9} & \scorecellblue{60.1} & \scorecellblue{59.6} & \scorecellblue{60.4} & \scorecellblue{51.4} & \scorecellblue{36.1}  & \scorecellblue{36.4} & - & - & \scorecellblue{65.9} \\

\midrule
 \multirow{6}{*}{L2} & Claude-3.5-Sonnet & \scorecellblue{40.0} & \scorecellblue{46.5} & \scorecellblue{44.3} & \scorecellblue{46.8} & \scorecellblue{43.7} & \scorecellblue{42.1} & \scorecellblue{38.1} & \scorecellblue{32.1} & \scorecellblue{24.7} & \scorecellblue{22.0} & \scorecellblue{47.8}\\
& GPT-4o & \scorecellblue{42.3} & \scorecellblue{40.7} & \scorecellblue{40.7} & \scorecellblue{41.5} & \scorecellblue{36.6} & \scorecellblue{37.0} & \scorecellblue{34.6} & \scorecellblue{30.4} & \scorecellblue{19.5} & \scorecellblue{22.9}  & \scorecellblue{40.0}  \\
& DeepSeek-V3 & \scorecellblue{41.2} & \scorecellblue{47.5} & \scorecellblue{48.0} & \scorecellblue{48.7} & \scorecellblue{44.7}  & \scorecellblue{45.3} & \scorecellblue{40.4} & \scorecellblue{35.5} & \scorecellblue{33.2} & \scorecellblue{35.3}   & \scorecellblue{54.0}\\
& Qwen2.5-Coder-7B & \scorecellblue{31.2} & \scorecellblue{23.0} & \scorecellblue{22.2} & \scorecellblue{19.7} & \scorecellblue{15.8} & \scorecellblue{13.3} & \scorecellblue{12.5} & \scorecellblue{7.4} & \scorecellblue{11.7} & \scorecellblue{7.2} & \scorecellblue{18.0} \\
& Qwen2.5-Coder-14B & \scorecellblue{35.5} & \scorecellblue{32.7} & \scorecellblue{32.0} & \scorecellblue{27.7} & \scorecellblue{29.2} & \scorecellblue{22.9} & \scorecellblue{23.5} & \scorecellblue{12.2} & \scorecellblue{14.4} & \scorecellblue{14.1}   & \scorecellblue{31.3} \\
& Qwen2.5-Coder-32B & \scorecellblue{34.7} & \scorecellblue{30.5} & \scorecellblue{23.5} & \scorecellblue{17.8} & \scorecellblue{15.1} & \scorecellblue{9.2} & \scorecellblue{7.0} & \scorecellblue{5.7} & \scorecellblue{4.1} & \scorecellblue{2.4}  & \scorecellblue{13.8}\\
\bottomrule
\end{tabular}
}\vspace{-0.3cm}
\end{table*}

Table~\ref{tab:RQ2_IA} reports the Instruction Accuracy (IA) results under the \textit{Dynamic Conversation} setting on the \bench dataset. Due to the requirement to incorporate feedback from the base task $I_1$, an additional round of interaction is required compared to \textit{Static Conversation}.
In L-1 tasks, most stronger LLMs exhibit strong adaptability, maintaining stable performance across interaction rounds. However, weaker LLM models such as the Qwen2.5-Coder series continue to exhibit unstable performance. This indicates that stronger LLMs can effectively utilize external feedback to correct code and reduce the negative impact of the number of rounds for following current instructions. However, in the L-2 dataset, the IA performance of most LLMs shows varying degrees of decline. For instance, Claude-3.5-Sonnet—one of the strongest closed-source models—achieved a score of merely 14.7 at Turn 6. 
This indicates that the long context may prevent the LLM from effectively utilizing feedback and benefiting from it. 

Table~\ref{tab:RQ2_CA} presents the Conversation Accuracy (CA) results. Same as \textit{Static Conversation}, we exclude ``Functionality Extension'' instructions in this experiment. 
In the L-1 task, similar to the \textit{Static Conversation} setting, all LLMs exhibit a trend of initially rising and then declining CA, indicating that even with feedback, their instruction-following ability at the session level gradually weakens as conversations progress. Compared with the feedback-free \textit{Static Conversation} results (final-round CA), models such as Claude and DeepSeek-V3 show notable improvements (68.0 → 74.5 and 57.4 → 72.3, respectively). This demonstrates that incorporating feedback effectively enhances LLMs’ instruction-following capability during interactive dialogues. Furthermore, the comparison between DeepSeek-V3 and Claude (74.5 vs. 72.3) suggests that feedback allows relatively weaker models to achieve performance comparable to stronger ones, thereby reducing potential deployment costs.
In the L-2 task, CIF also achieves substantial gains over the \textit{Static Conversation} setting without feedback (e.g., Claude: 25.8 → 47.8; DeepSeek-V3: 25.6 → 40.0), further highlighting feedback’s role in strengthening instruction following. However, CA still declines notably as conversations lengthen, indicating that LLMs continue to face challenges in maintaining consistent instruction following across extended multi-turn interactions.

\begin{figure}[h]
    \centering

    \begin{subfigure}[b]{0.48\linewidth}
        \includegraphics[width=\linewidth]{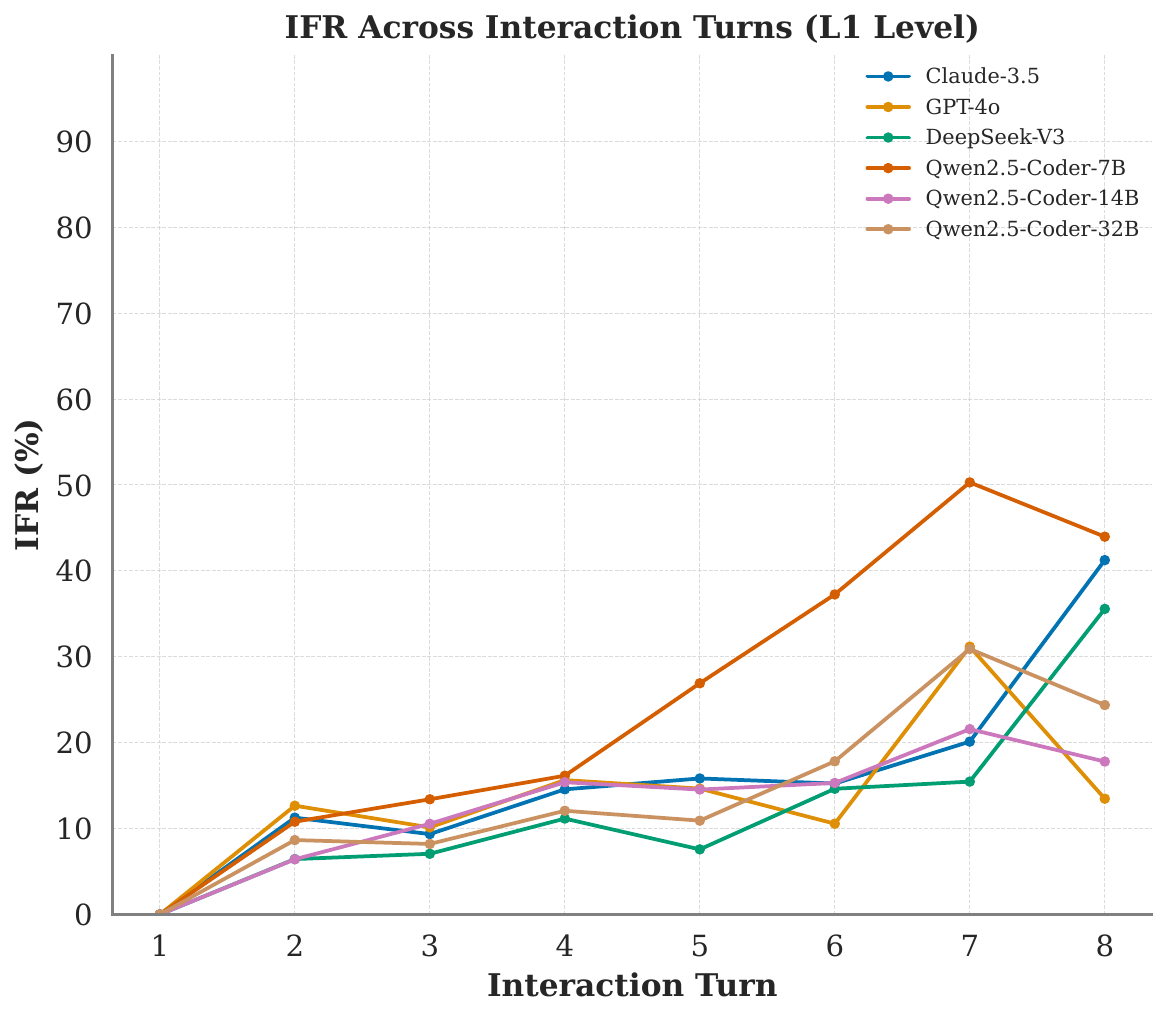}
        \caption{L1-Dynamic Conversation}
        \label{fig:L1}
    \end{subfigure}
    \hfill
    \begin{subfigure}[b]{0.48\linewidth}
        \includegraphics[width=\linewidth]{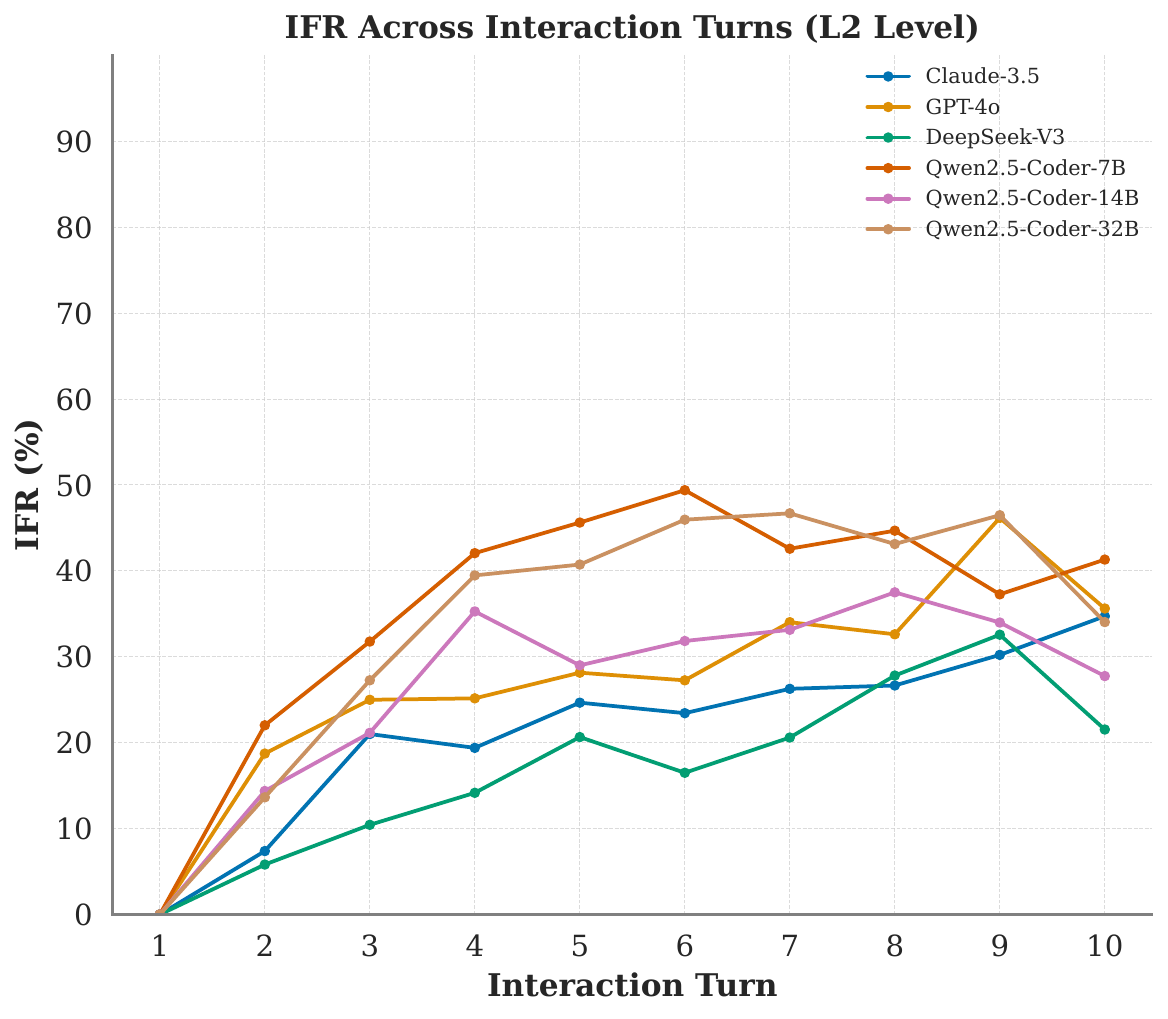}
        \caption{L2-Dynamic Conversation}
        \label{fig:dev_turn}
    \end{subfigure}

    \caption{IFR results in Dynamic Conversation}
    \label{fig:RQ2_IFR}
\end{figure}

We further analysed the Instruction Forgetting Rate (IFR), as illustrated in Figure \ref{fig:RQ2_IFR}. Across both L-1 and L-2 tasks, IFR progressively increases with dialogue depth, suggesting that LLMs struggle to maintain conversational state when handling multiple instructions simultaneously. During such exchanges, models often lose control over editing operations, inadvertently modifying previously correct content and exhibiting forgetting behaviour. Interestingly, in both L-1 and L-2 settings, a subset of IFR trajectories shows an initial rise followed by a decline, corresponding to dialogue rounds where users restate previously forgotten instructions. This finding indicates that enabling LLMs to recognise and recover forgotten or previously executed instructions can potentially improve their instruction-following performance in dynamic interaction scenarios.

\begin{RQbox} \textbf{RQ2 Summary: }
In the \textit{Dynamic Conversation} setting, stronger LLMs effectively leverage feedback to maintain stable instruction-following performance in L-1 tasks. However, in L-2 tasks, extended context severely hinders feedback utilization, leading to declines in both IA and CA. Although feedback improves instruction following numbers during conversations, performance still degrades over longer interactions as Instruction Forgetting Rate (IFR) increases. These results highlight that enabling LLMs to recognise and recover forgotten or previously executed instructions is a potential factor for enhancing robustness in long-context, multi-turn scenarios.
\end{RQbox}
\subsection{RQ3: Performance on Various VI Strategies}
\label{RQ3}
\begin{figure}[h]
    \centering

    \begin{subfigure}[b]{0.48\linewidth}
        \includegraphics[width=\linewidth]{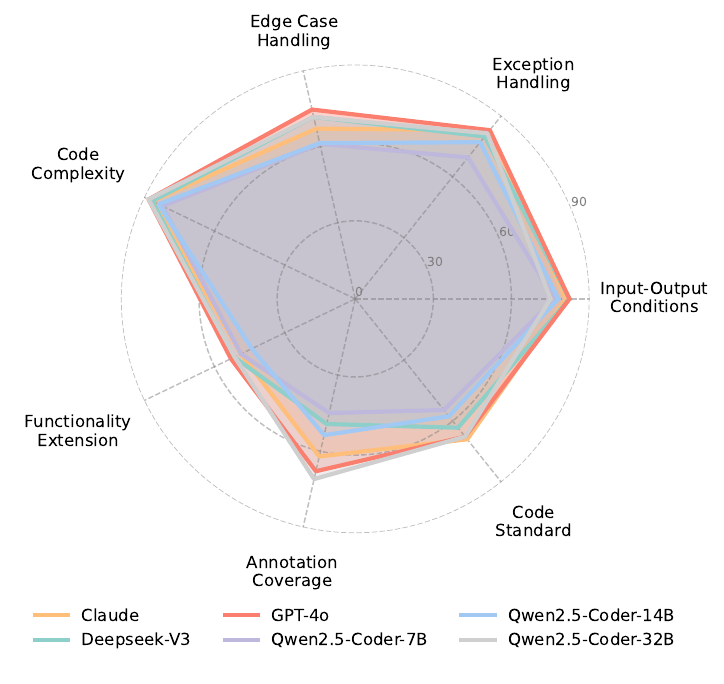}
        \caption{L1}
        \label{fig:L1}
    \end{subfigure}
    \hfill
    \begin{subfigure}[b]{0.48\linewidth}
        \includegraphics[width=\linewidth]{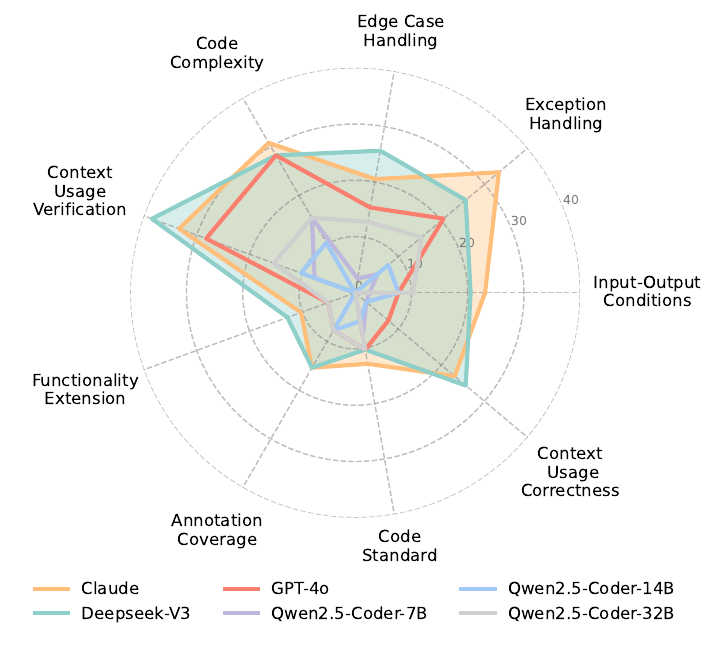}
        \caption{L2}
        \label{fig:dev_turn}
    \end{subfigure}

    \caption{The IA results on various VI strategies}
    \label{fig:RQ3}
\end{figure}

To evaluate LLMs' instruction following performance across different instruction types, we implemented a 2-round experimental design: the first round comprised basic programming tasks, while the second round focused on verifiable instructions. This design aims to controlle for potential confounding effects from varying dialogue round counts. As illustrated in Figure \ref{fig:RQ3}, the L-1 dataset revealed significant performance variations in IA scores across instruction types, particularly in ``Annotation Coverage'', ``Code Standard'', and ``Functionality Extension''. We also find that the underperforming Turns 5, 7, and 8 in Table \ref{tab:IA_RQ1} mostly correspond to these three instruction strategies. This suggests that LLMs may have insufficient training data for non-functional requirements (\textit{e.g.}, annotations and code style) and functional enhancement scenarios, resulting in comparatively weaker performance. 

The results in the L-2 task further demonstrate substantial performance disparities between different LLMs and instruction types. Overall, the performance of LLM declined compared to L-1, further indicating that the long context of L-2 is an important factor affecting the instruction following ability of LLM. Closed-source models (\textit{e.g.}, Claude) and large-parameter open-source models (\textit{e.g.}, DeepSeek-V3) substantially outperformed small-parameter SOTA open-source models like Qwen2.5-Coder, particularly in development scenarios requiring longer context retention. Consistent with L-1 findings, instructions such as ``Annotation Coverage'' again showed relatively poor performance, reinforcing the observed limitations of LLMs in handling such requirements.

Notably, LLMs exhibited strong performance in ``Context Usage Verification'', indicating their capability to identify relevant contextual information while demonstrating poor performance in and ``Context Usage Correctness''. This dichotomy suggests that while LLMs can effectively recognize correct contextual information, their ability to appropriately apply this context requires further enhancement.

\begin{RQbox} \textbf{RQ3 Summary: }
LLMs demonstrate suboptimal performance on instruction strategies such as ``Annotation Coverage'', ``Code Standard'', and ``Functionality Extension'', likely due to insufficient training in these domains. Additionally, longer context lengths negatively impact the performance of LLM across all instruction types. While LLMs can effectively identify relevant information from context, their ability to utilize it accurately remains limited.
\end{RQbox}

\subsection{RQ4: Exploration of IF Enhancement}
To further enhance LLMs’ instruction following performance in interactive code generation, one intuitive and lightweight approach is to improve the prompting strategy. Unlike model fine-tuning or reinforcement optimization, which require extensive data and computational resources, prompt enhancement focuses on optimizing the interaction layer between the developer and the model, making it more practical. Therefore, we explore whether carefully designed prompting strategies can improve LLMs’ ability to interpret and execute evolving developer instructions across multiple turns.
To investigate this question, we examine three representative prompting strategies that capture different aspects of context management and reasoning: 
\begin{itemize}
    \item \textbf{Full History (FH)}: preserves all prior user–LLM interactions as input for the next round (consistent with \textit{Static Conversation}). This retains all prior user–LLM interactions as input for the next round, enabling the model to leverage complete conversational context.
    \item \textbf{Chain-of-Thought (CoT)}\citep{cot}: instructs the LLM to explicitly generate a reasoning process before producing the final answer in each round. The objective is to investigate whether explicit reasoning enhances the model’s instruction following performance in multi-turn scenarios.
    \item \textbf{Cumulative Instruction (CI)}: retains only the user instructions from all previous rounds, excluding model responses, as context for subsequent turns. For example, in $i$-th round, compared to FH's prompt $(I_1, A_1, I_2, A_2, ...,I_i)$, the prompt is $(I_1,I_2,...,I_i)$. This approach highlights the trajectory of developer intent while mitigating the accumulation of errors introduced by earlier model outputs, allowing assessing whether selective context preservation improves overall performance.
\end{itemize}

The Instruction Accuracy (IA) results of the three prompt-enhancing methods are presented in Table \ref{tab:IA_RQ4}. As shown in the results, in the L-1 task, consistent with single-turn performance, the CoT method improves adherence to current instructions through explicit reasoning, whereas CI does not produce significant improvement. In contrast, in the L-2 task, CI achieves the best performance, while CoT remains effective but suboptimal. These findings indicate that accumulated conversational context can degrade LLM performance, while CoT mitigates this degradation through structured reasoning. However, retaining only essential user instructions (as in CI) further enhances multi-turn instruction following.

\begin{table*}[t]
\centering
\small
\setlength{\abovecaptionskip}{0.1cm}
\caption{The IA results for different prompting methods. The backbone model is DeepSeek-V3}
\label{tab:IA_RQ4}
\resizebox{\textwidth}{!}{
\begin{tabular}{l|l|ccccccccccc}
\toprule
\textbf{Level} & \textbf{Model} & Turn-1 & Turn-2 & Turn-3 & Turn-4 & Turn-5 & Turn-6 & Turn-7 & Turn-8 & Turn-9 & Turn-10 & Avg. \\
\midrule
 \multirow{3}{*}{L1} 
& FH & \scorecellgreen{60.0} & \scorecellgreen{80.0} & \scorecellgreen{84.3} & \scorecellgreen{77.4} & \scorecellgreen{55.7} & \scorecellgreen{71.8} & \scorecellgreen{29.1} & \scorecellgreen{54.9} & - & - & \scorecellgreen{64.2} \\
& CoT
& \scorecellgreen{57.1} 
& \scorecellgreen{78.6} 
& \scorecellgreen{84.3} 
& \scorecellgreen{74.3} 
& \scorecellgreen{60.0} 
& \scorecellgreen{70.4}
& \scorecellgreen{43.5} 
& \scorecellgreen{53.6} & - & - 
& \scorecellgreen{65.2} \\
& CI 
& \scorecellgreen{58.6} 
& \scorecellgreen{81.4}
& \scorecellgreen{85.7} 
& \scorecellgreen{72.9} 
& \scorecellgreen{55.7} 
& \scorecellgreen{77.7} 
& \scorecellgreen{24.8}
& \scorecellgreen{58.3} & - & - 
& \scorecellgreen{64.4} \\
\midrule
 \multirow{3}{*}{L2} 
& FH & \scorecellgreen{50.0} & \scorecellgreen{55.0} & \scorecellgreen{62.5} & \scorecellgreen{52.5} & \scorecellgreen{27.5} & \scorecellgreen{30.0} & \scorecellgreen{2.5} & \scorecellgreen{45.5} & \scorecellgreen{45.5} & \scorecellgreen{27.5} & \scorecellgreen{39.8} \\

& CoT
& \scorecellgreen{55.0} 
& \scorecellgreen{62.5} 
& \scorecellgreen{60.0} 
& \scorecellgreen{55.0} 
& \scorecellgreen{37.5} 
& \scorecellgreen{37.5}
& \scorecellgreen{10.0} 
& \scorecellgreen{55.0} 
& \scorecellgreen{47.5} 
& \scorecellgreen{35.0} 
& \scorecellgreen{45.5} \\
& CI 
& \scorecellgreen{52.5} 
& \scorecellgreen{62.5}
& \scorecellgreen{57.5} 
& \scorecellgreen{70.0} 
& \scorecellgreen{47.5} 
& \scorecellgreen{27.5} 
& \scorecellgreen{5.0}
& \scorecellgreen{70.0} 
& \scorecellgreen{55.0} 
& \scorecellgreen{40.0} 
& \scorecellgreen{48.8} \\

\bottomrule
\end{tabular}
}
\vspace{-0.3cm}
\end{table*}

\begin{table*}[t]
\centering
\small
\setlength{\abovecaptionskip}{0.1cm}
\caption{The CA results for different prompting methods. The backbone model is DeepSeek-V3. \textit{Tokens} represents the average total number of tokens required per conversation}
\label{tab:CA_RQ4}
\resizebox{\textwidth}{!}{
\begin{tabular}{l|l|ccccccccccc}
\toprule
\textbf{Level} & \textbf{Model} & Turn-1 & Turn-2 & Turn-3 & Turn-4 & Turn-5 & Turn-6 & Turn-7 & Turn-8 & Turn-9 & Avg. & Tokens. \\
\midrule
 \multirow{3}{*}{L1} 
& FH 
 & \scorecellblue{60.0}
 & \scorecellblue{77.6}
 & \scorecellblue{78.0}
 & \scorecellblue{77.4}
 & \scorecellblue{72.8}
 & \scorecellblue{62.4}
& \scorecellblue{57.4}
& -& -
& \scorecellblue{69.4} & 8.2K\\
& CoT 
 & \scorecellblue{57.1}
& \scorecellblue{75.4}
 & \scorecellblue{76.4}
 & \scorecellblue{75.4}
& \scorecellblue{72.0}
 & \scorecellblue{65.7}
 & \scorecellblue{63.1}
 & -& -
 & \scorecellblue{69.3} & 21.7K\\
& CI
 & \scorecellblue{58.6}
 & \scorecellblue{78.5}
& \scorecellblue{78.6}
 & \scorecellblue{78.3}
 & \scorecellblue{73.3}
 & \scorecellblue{71.9}
& \scorecellblue{67.7}
 & -& -
 & \scorecellblue{72.4} & 4.2K\\

\midrule
 \multirow{3}{*}{L2} 
& FH & \scorecellblue{50.0} & \scorecellblue{51.2} & \scorecellblue{54.2} & \scorecellblue{51.2} & \scorecellblue{35.0} & \scorecellblue{32.1} & \scorecellblue{29.3} & \scorecellblue{31.6} & \scorecellblue{33.6}& \scorecellblue{40.9} & 71.3K\\

&  CoT
& \scorecellblue{55.0} 
& \scorecellblue{57.5} 
& \scorecellblue{55.8} 
& \scorecellblue{55.7} 
& \scorecellblue{46.0} 
& \scorecellblue{42.5}
& \scorecellblue{37.1} 
& \scorecellblue{41.3} 
& \scorecellblue{41.1} 
& \scorecellblue{48.0} & 79.1K\\
& CI 
& \scorecellblue{52.5} 
& \scorecellblue{57.5}
& \scorecellblue{59.2} 
& \scorecellblue{60.6} 
& \scorecellblue{54.0} 
& \scorecellblue{50.4} 
& \scorecellblue{45.4}
& \scorecellblue{46.9} 
& \scorecellblue{47.8} 
& \scorecellblue{52.7} & 54.3K\\

\bottomrule
\end{tabular}
}
\vspace{-0.3cm}
\end{table*}

Table \ref{tab:CA_RQ4} presents the Conversational Accuracy (CA) results as well as the computational costs (Tokens). These results further demonstrate that CI not only achieves the highest CA in both L-1 and L-2 tasks but also incurs the lowest computational overhead. This underscores the promise of context compression—extracting key elements from dialogue history—to maintain both efficiency and instruction following.
Future work could explore adaptive context selection mechanisms that dynamically identify and preserve critical historical information, thereby enabling LLMs to enhance conversational performance.
\begin{RQbox} \textbf{RQ4 Summary: }
Excessive conversational history can hinder instruction following, but reasoning-based prompting (CoT) and cumulative instruction (CI) alleviate this issue from different perspectives. CI, in particular, offers a more efficient and scalable approach for sustaining LLM performance across multi-turn interactions. Exploring to preserve critical historical information is a potential method for efficiently enhancing LLMs’ conversational performance.
\end{RQbox}

\subsection{Case Study}
\begin{figure*}[h]
    \centering
    \includegraphics[width=\linewidth]{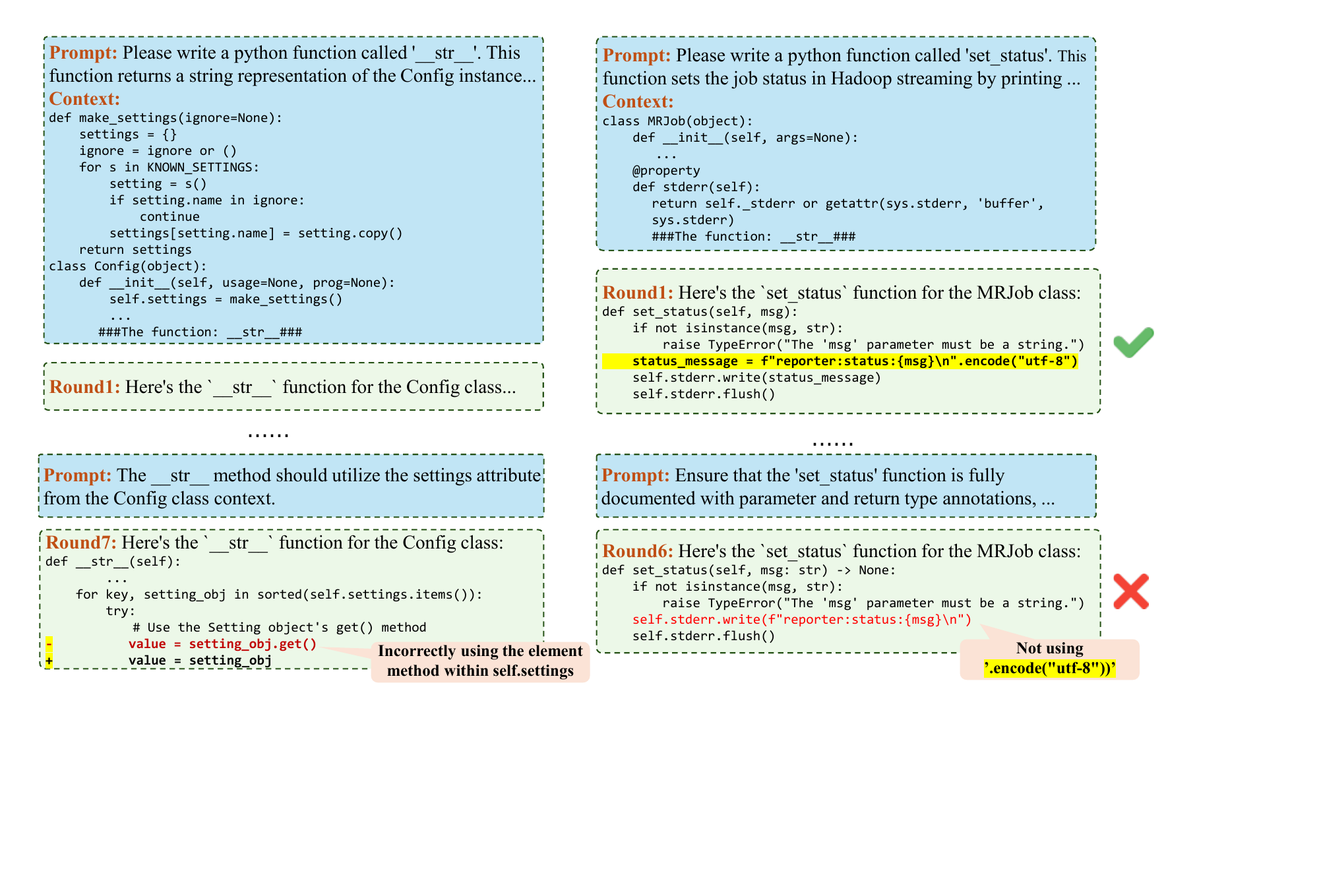}
	\caption{Case analysis of DeepSeek-V3 in L-2 Tasks}
	\label{fig:case_study}
    \vspace{-0.5cm}
\end{figure*}
To provide a more intuitive illustration of LLM performance on CodeIF-Bench, Figure \ref{fig:case_study} presents representative examples of DeepSeek-V3’s responses in L-2 tasks. In the left-hand example, during the 7-th interaction round, the model was instructed to use a class member variable within the given context. While it successfully identified and referenced the variable, it mistakenly invoked a non-existent method. This suggests that although LLMs can recognize relevant contextual information, their ability to apply it accurately remains limited. In real-world development scenarios, LLMs must not only retrieve relevant repository context but also comprehend and apply it correctly to follow user instructions—highlighting a key challenge in repository-level code generation.
The right-hand example shows that, as conversational context accumulates over multiple rounds, the model erroneously modifies a previously correct line of code in the sixth round. This demonstrates that cumulative dialogue context can degrade an LLM’s instruction-following ability. Hence, enabling LLMs to extract effective information from accumulated dialogue and recognize forgotten or previously executed instructions, represents a possible direction for enhancing their robustness in multi-turn interactions.

In summary, within real-world repository development, LLMs must effectively manage two intertwined forms of context—repository context and conversational context. The ability to integrate and reason over both is crucial for improving instruction-following performance.

\section{Threats to Validity}
\textbf{Threats to External Validity} arise from potential data leakage. While MBPP and DevEval are widely adopted benchmarks, they may have been exposed in public datasets used for model training. However, the verifiable instructions except for instructions from MBPP and DevEval in \bench were meticulously manually annotated and not sourced from public corpora to mitigate leakage risks. Moreover, our experimental results demonstrate substantial performance variations across all evaluated LLMs on \bench, with no observable fitting bias, suggesting that data leakage has minimal impact on evaluation accuracy. To further address this concern, we will implement periodic dataset updates to manage potential leakage risks proactively.

\textbf{Threats to Internal Validity} stem from the fact that \bench is a single-language dataset focused solely on Python, neglecting other languages. However, our method is language-agnostic, and in future work, we will incorporate other programming languages such as Java. Additionally, the validity threat also arises from the strategies of instruction. Although we collected them from real review comments, they may not cover all instruction strategies. In future work, we will regularly update the instruction strategies and data to address this threat.

\textbf{Threats to Construct Validity} relate to manual annotation. Although we hired experienced programmers for annotation and reached consensus on all data through discussion, potential defects and errors may still exist. We have open-sourced the data to collect any errors that may arise and update our dataset on time.

\section{Conclusion}
This paper presents \bench, the first benchmark for evaluating LLMs' instruction-following capabilities in interactive code generation. \bench incorporates verifiable instructions aligned with the real-world software development, complemented by corresponding test cases, enabling a comprehensive evaluation of LLMs' instruction-following performance in multi-turn code generation scenarios. 
In both \textit{Static Conversation} and \textit{Dynamic Conversation} settings, we evaluate the performance of 6 state-of-the-art LLMs and summarize important influencing factors and possible areas for improvement. In our future research, we will expand \bench by incorporating additional programming languages and exploring methods to enhance the interactive code generation capabilities of LLMs.

\section*{Declarations}
\subsection*{Funding}
This research is supported by the National Natural Science Foundation of China Grants Nos. 62302021 and 62177003.
\subsection*{Author Contributions}
\textbf{Peiding Wang:} Data curation, Methodology, Experiments, Writing-original draft; 
\textbf{Li Zhang:} Methodology, Writing-review \& editing;
\textbf{Fang Liu:} Data curation, Methodology, Writing-review \& editing; 
\textbf{Lin Shi:} Methodology, Writing-review \& editing; 
\textbf{Minxiao Li:} Data curation, Methodology, Experiments;  
\textbf{Bo Shen:} Writing-review \& editing; 
\textbf{An Fu:} Writing-review \& editing.


\subsection*{Data Availability} The dataset and the source code are available at \url{https://github.com/zhu-zhu-ding/CodeIF-Bench}.

\subsection*{Conflict of Interest}
The authors have no competing interests to declare that are relevant to the content of this paper.

\subsection*{Ethical Approval}
Not applicable to this study.

\subsection*{Informed Consent}
Not applicable to this study.

\bibliography{sn-bibliography}

@article{zhu2024deepseek,
  title={DeepSeek-Coder-V2: Breaking the Barrier of Closed-Source Models in Code Intelligence},
  author={Zhu, Qihao and Guo, Daya and Shao, Zhihong and Yang, Dejian and Wang, Peiyi and Xu, Runxin and Wu, Y and Li, Yukun and Gao, Huazuo and Ma, Shirong and others},
  journal={arXiv preprint arXiv:2406.11931},
  year={2024}
}

@article{roziere2023code,
  title={Code llama: Open foundation models for code},
  author={Roziere, Baptiste and Gehring, Jonas and Gloeckle, Fabian and Sootla, Sten and Gat, Itai and Tan, Xiaoqing Ellen and Adi, Yossi and Liu, Jingyu and Sauvestre, Romain and Remez, Tal and others},
  journal={arXiv preprint arXiv:2308.12950},
  year={2023}
}

@inproceedings{mt-bench-101,
    title = "{MT}-Bench-101: A Fine-Grained Benchmark for Evaluating Large Language Models in Multi-Turn Dialogues",
    author = "Bai, Ge  and
      Liu, Jie  and
      Bu, Xingyuan  and
      He, Yancheng  and
      Liu, Jiaheng  and
      Zhou, Zhanhui  and
      Lin, Zhuoran  and
      Su, Wenbo  and
      Ge, Tiezheng  and
      Zheng, Bo  and
      Ouyang, Wanli",
    editor = "Ku, Lun-Wei  and
      Martins, Andre  and
      Srikumar, Vivek",
    booktitle = "Proceedings of the 62nd Annual Meeting of the Association for Computational Linguistics (Volume 1: Long Papers)",
    month = aug,
    year = "2024",
    address = "Bangkok, Thailand",
    publisher = "Association for Computational Linguistics",
    url = "https://aclanthology.org/2024.acl-long.401/",
    doi = "10.18653/v1/2024.acl-long.401",
    pages = "7421--7454"
}

@misc{codeif,
      title={CodeIF: Benchmarking the Instruction-Following Capabilities of Large Language Models for Code Generation}, 
      author={Kaiwen Yan and Hongcheng Guo and Xuanqing Shi and Jingyi Xu and Yaonan Gu and Zhoujun Li},
      year={2025},
      eprint={2502.19166},
      archivePrefix={arXiv},
      primaryClass={cs.SE},
      url={https://arxiv.org/abs/2502.19166}, 
}

@misc{multiif,
      title={Multi-IF: Benchmarking LLMs on Multi-Turn and Multilingual Instructions Following}, 
      author={Yun He and Di Jin and Chaoqi Wang and Chloe Bi and Karishma Mandyam and Hejia Zhang and Chen Zhu and Ning Li and Tengyu Xu and Hongjiang Lv and Shruti Bhosale and Chenguang Zhu and Karthik Abinav Sankararaman and Eryk Helenowski and Melanie Kambadur and Aditya Tayade and Hao Ma and Han Fang and Sinong Wang},
      year={2024},
      eprint={2410.15553},
      archivePrefix={arXiv},
      primaryClass={cs.CL},
      url={https://arxiv.org/abs/2410.15553}, 
}

@article{luo2023wizardcoder,
  title={Wizardcoder: Empowering code large language models with evol-instruct},
  author={Luo, Ziyang and Xu, Can and Zhao, Pu and Sun, Qingfeng and Geng, Xiubo and Hu, Wenxiang and Tao, Chongyang and Ma, Jing and Lin, Qingwei and Jiang, Daxin},
  journal={arXiv preprint arXiv:2306.08568},
  year={2023}
}

@misc{GitHub-Copilot,
  author = {Github},
  title = {Github Copilot},
  year = {2021},
  url = {https://github.com/features/copilot}
}

@misc{Cursor,
  author = {Anysphere},
  title = {Cursor},
  year = {2023},
  url = {https://www.cursor.com/}
}

@article{HumanEval,
  title={Evaluating large language models trained on code},
  author={Chen, Mark and Tworek, Jerry and Jun, Heewoo and Yuan, Qiming and Pinto, Henrique Ponde De Oliveira and Kaplan, Jared and Edwards, Harri and Burda, Yuri and Joseph, Nicholas and Brockman, Greg and others},
  journal={arXiv preprint arXiv:2107.03374},
  year={2021}
}

@article{MBPP,
  title={Program synthesis with large language models},
  author={Austin, Jacob and Odena, Augustus and Nye, Maxwell and Bosma, Maarten and Michalewski, Henryk and Dohan, David and Jiang, Ellen and Cai, Carrie and Terry, Michael and Le, Quoc and others},
  journal={arXiv preprint arXiv:2108.07732},
  year={2021}
}

@inproceedings{CoderEval,
  title={Codereval: A benchmark of pragmatic code generation with generative pre-trained models},
  author={Yu, Hao and Shen, Bo and Ran, Dezhi and Zhang, Jiaxin and Zhang, Qi and Ma, Yuchi and Liang, Guangtai and Li, Ying and Wang, Qianxiang and Xie, Tao},
  booktitle={Proceedings of the 46th IEEE/ACM International Conference on Software Engineering},
  pages={1--12},
  year={2024}
}

@article{MT-Bench,
  title={Judging llm-as-a-judge with mt-bench and chatbot arena},
  author={Zheng, Lianmin and Chiang, Wei-Lin and Sheng, Ying and Zhuang, Siyuan and Wu, Zhanghao and Zhuang, Yonghao and Lin, Zi and Li, Zhuohan and Li, Dacheng and Xing, Eric and others},
  journal={Advances in Neural Information Processing Systems},
  volume={36},
  pages={46595--46623},
  year={2023}
}

@article{MT-Eval,
  title={MT-Eval: A Multi-Turn Capabilities Evaluation Benchmark for Large Language Models},
  author={Kwan, Wai-Chung and Zeng, Xingshan and Jiang, Yuxin and Wang, Yufei and Li, Liangyou and Shang, Lifeng and Jiang, Xin and Liu, Qun and Wong, Kam-Fai},
  journal={arXiv preprint arXiv:2401.16745},
  year={2024}
}

@misc{FB-Bench,
      title={FB-Bench: A Fine-Grained Multi-Task Benchmark for Evaluating LLMs' Responsiveness to Human Feedback}, 
      author={Youquan Li and Miao Zheng and Fan Yang and Guosheng Dong and Bin Cui and Weipeng Chen and Zenan Zhou and Wentao Zhang},
      year={2024},
      eprint={2410.09412},
      archivePrefix={arXiv},
      primaryClass={cs.CL},
      url={https://arxiv.org/abs/2410.09412}, 
}

@article{IF-Bench,
  title={Instruction-following evaluation for large language models},
  author={Zhou, Jeffrey and Lu, Tianjian and Mishra, Swaroop and Brahma, Siddhartha and Basu, Sujoy and Luan, Yi and Zhou, Denny and Hou, Le},
  journal={arXiv preprint arXiv:2311.07911},
  year={2023}
}

@article{Systematic,
  title={Systematic evaluation of llm-as-a-judge in llm alignment tasks: Explainable metrics and diverse prompt templates},
  author={Wei, Hui and He, Shenghua and Xia, Tian and Wong, Andy and Lin, Jingyang and Han, Mei},
  journal={arXiv preprint arXiv:2408.13006},
  year={2024}
}

@article{huang2024empirical,
  title={An empirical study of llm-as-a-judge for llm evaluation: Fine-tuned judge models are task-specific classifiers},
  author={Huang, Hui and Qu, Yingqi and Liu, Jing and Yang, Muyun and Zhao, Tiejun},
  journal={arXiv preprint arXiv:2403.02839},
  year={2024}
}

@article{chen2024mllm,
  title={Mllm-as-a-judge: Assessing multimodal llm-as-a-judge with vision-language benchmark},
  author={Chen, Dongping and Chen, Ruoxi and Zhang, Shilin and Liu, Yinuo and Wang, Yaochen and Zhou, Huichi and Zhang, Qihui and Wan, Yao and Zhou, Pan and Sun, Lichao},
  journal={arXiv preprint arXiv:2402.04788},
  year={2024}
}

@inproceedings{DevEval,
  author       = {Jia Li and
                  Ge Li and
                  Yunfei Zhao and
                  Yongmin Li and
                  Huanyu Liu and
                  Hao Zhu and
                  Lecheng Wang and
                  Kaibo Liu and
                  Zheng Fang and
                  Lanshen Wang and
                  Jiazheng Ding and
                  Xuanming Zhang and
                  Yuqi Zhu and
                  Yihong Dong and
                  Zhi Jin and
                  Binhua Li and
                  Fei Huang and
                  Yongbin Li and
                  Bin Gu and
                  Mengfei Yang},
  title        = {DevEval: {A} Manually-Annotated Code Generation Benchmark Aligned
                  with Real-World Code Repositories},
  booktitle    = {{ACL} (Findings)},
  pages        = {3603--3614},
  publisher    = {Association for Computational Linguistics},
  year         = {2024}
}

@article{CanItEdit,
  title={Can it edit? evaluating the ability of large language models to follow code editing instructions},
  author={Cassano, Federico and Li, Luisa and Sethi, Akul and Shinn, Noah and Brennan-Jones, Abby and Ginesin, Jacob and Berman, Edward and Chakhnashvili, George and Lozhkov, Anton and Anderson, Carolyn Jane and others},
  journal={arXiv preprint arXiv:2312.12450},
  year={2023}
}

@article{RepoEval,
  title={Repocoder: Repository-level code completion through iterative retrieval and generation},
  author={Zhang, Fengji and Chen, Bei and Zhang, Yue and Keung, Jacky and Liu, Jin and Zan, Daoguang and Mao, Yi and Lou, Jian-Guang and Chen, Weizhu},
  journal={arXiv preprint arXiv:2303.12570},
  year={2023}
}

@misc{InfoBench,
      title={InFoBench: Evaluating Instruction Following Ability in Large Language Models}, 
      author={Yiwei Qin and Kaiqiang Song and Yebowen Hu and Wenlin Yao and Sangwoo Cho and Xiaoyang Wang and Xuansheng Wu and Fei Liu and Pengfei Liu and Dong Yu},
      year={2024},
      eprint={2401.03601},
      archivePrefix={arXiv},
      primaryClass={cs.CL},
      url={https://arxiv.org/abs/2401.03601}, 
}

@misc{codereview,
      title={Automating Code Review Activities by Large-Scale Pre-training}, 
      author={Zhiyu Li and Shuai Lu and Daya Guo and Nan Duan and Shailesh Jannu and Grant Jenks and Deep Majumder and Jared Green and Alexey Svyatkovskiy and Shengyu Fu and Neel Sundaresan},
      year={2022},
      eprint={2203.09095},
      archivePrefix={arXiv},
      primaryClass={cs.SE},
      url={https://arxiv.org/abs/2203.09095}, 
}

@inproceedings{perry2023users,
  title={Do users write more insecure code with AI assistants?},
  author={Perry, Neil and Srivastava, Megha and Kumar, Deepak and Boneh, Dan},
  booktitle={Proceedings of the 2023 ACM SIGSAC Conference on Computer and Communications Security},
  pages={2785--2799},
  year={2023}
}

@misc{qwencoder,
      title={Qwen2.5-Coder Technical Report}, 
      author={Binyuan Hui and Jian Yang and Zeyu Cui and Jiaxi Yang and Dayiheng Liu and Lei Zhang and Tianyu Liu and Jiajun Zhang and Bowen Yu and Keming Lu and Kai Dang and Yang Fan and Yichang Zhang and An Yang and Rui Men and Fei Huang and Bo Zheng and Yibo Miao and Shanghaoran Quan and Yunlong Feng and Xingzhang Ren and Xuancheng Ren and Jingren Zhou and Junyang Lin},
      year={2024},
      eprint={2409.12186},
      archivePrefix={arXiv},
      primaryClass={cs.CL},
      url={https://arxiv.org/abs/2409.12186}, 
}

@article{openai2024gpt,
  title={Gpt-4 technical report, 2024},
  author={OpenAI, Josh Achiam and Adler, Steven and Agarwal, Sandhini and Ahmad, Lama and Akkaya, Ilge and Aleman, Florencia Leoni and Almeida, Diogo and Altenschmidt, Janko and Altman, Sam and Anadkat, Shyamal and others},
  journal={URL https://arxiv. org/abs/2303.08774},
  volume={2},
  pages={6},
  year={2024}
}

@misc{claude,
  author = {claude},
  title = {Claude},
  year = {2023},
  url = {https://claude.ai/}
}

@inproceedings{APPS,
  author       = {Dan Hendrycks and
                  Steven Basart and
                  Saurav Kadavath and
                  Mantas Mazeika and
                  Akul Arora and
                  Ethan Guo and
                  Collin Burns and
                  Samir Puranik and
                  Horace He and
                  Dawn Song and
                  Jacob Steinhardt},
  editor       = {Joaquin Vanschoren and
                  Sai{-}Kit Yeung},
  title        = {Measuring Coding Challenge Competence With {APPS}},
  booktitle    = {Proceedings of the Neural Information Processing Systems Track on
                  Datasets and Benchmarks 1, NeurIPS Datasets and Benchmarks 2021, December
                  2021, virtual},
  year         = {2021},
  url          = {https://datasets-benchmarks-proceedings.neurips.cc/paper/2021/hash/c24cd76e1ce41366a4bbe8a49b02a028-Abstract-round2.html},
  bibsource    = {dblp computer science bibliography, https://dblp.org}
}

@article{ClassEval,
  author       = {Xueying Du and
                  Mingwei Liu and
                  Kaixin Wang and
                  Hanlin Wang and
                  Junwei Liu and
                  Yixuan Chen and
                  Jiayi Feng and
                  Chaofeng Sha and
                  Xin Peng and
                  Yiling Lou},
  title        = {ClassEval: {A} Manually-Crafted Benchmark for Evaluating LLMs on Class-level
                  Code Generation},
  journal      = {CoRR},
  volume       = {abs/2308.01861},
  year         = {2023},
  url          = {https://doi.org/10.48550/arXiv.2308.01861},
  doi          = {10.48550/arXiv.2308.01861},
  eprinttype    = {arXiv},
  eprint       = {2308.01861},
  bibsource    = {dblp computer science bibliography, https://dblp.org}
}

@misc{codebugs,
      title={Bugs in Large Language Models Generated Code: An Empirical Study}, 
      author={Florian Tambon and Arghavan Moradi Dakhel and Amin Nikanjam and Foutse Khomh and Michel C. Desmarais and Giuliano Antoniol},
      year={2024},
      eprint={2403.08937},
      archivePrefix={arXiv},
      primaryClass={cs.SE},
      url={https://arxiv.org/abs/2403.08937}, 
}

@inproceedings{codeweaknesses,
  title={Weaknesses in LLM-Generated Code for Embedded Systems Networking},
  author={Dunne, Murray and Schram, Kylee and Fischmeister, Sebastian},
  booktitle={2024 IEEE 24th International Conference on Software Quality, Reliability and Security (QRS)},
  pages={250--261},
  year={2024},
  organization={IEEE}
}

@inproceedings{kcenter,
  title={Active Learning for Convolutional Neural Networks: A Core-Set Approach},
  author={Sener, Ozan and Savarese, Silvio},
  booktitle={International Conference on Learning Representations},
  year={2018}
}

@inproceedings{chen-etal-2024-humans,
    title = "Humans or {LLM}s as the Judge? A Study on Judgement Bias",
    author = "Chen, Guiming Hardy  and
      Chen, Shunian  and
      Liu, Ziche  and
      Jiang, Feng  and
      Wang, Benyou",
    editor = "Al-Onaizan, Yaser  and
      Bansal, Mohit  and
      Chen, Yun-Nung",
    booktitle = "Proceedings of the 2024 Conference on Empirical Methods in Natural Language Processing",
    month = nov,
    year = "2024",
    address = "Miami, Florida, USA",
    publisher = "Association for Computational Linguistics",
    url = "https://aclanthology.org/2024.emnlp-main.474/",
    doi = "10.18653/v1/2024.emnlp-main.474",
    pages = "8301--8327"
}

@misc{li2025preferenceleakagecontaminationproblem,
      title={Preference Leakage: A Contamination Problem in LLM-as-a-judge}, 
      author={Dawei Li and Renliang Sun and Yue Huang and Ming Zhong and Bohan Jiang and Jiawei Han and Xiangliang Zhang and Wei Wang and Huan Liu},
      year={2025},
      eprint={2502.01534},
      archivePrefix={arXiv},
      primaryClass={cs.LG},
      url={https://arxiv.org/abs/2502.01534}, 
}

@article{ou2024inductive,
  title={Inductive-Deductive Strategy Reuse for Multi-Turn Instructional Dialogues},
  author={Ou, Jiao and Wu, Jiayu and Liu, Che and Zhang, Fuzheng and Zhang, Di and Gai, Kun},
  journal={arXiv preprint arXiv:2404.11095},
  year={2024}
}

@misc{cot,
      title={Large Language Models are Zero-Shot Reasoners}, 
      author={Takeshi Kojima and Shixiang Shane Gu and Machel Reid and Yutaka Matsuo and Yusuke Iwasawa},
      year={2023},
      eprint={2205.11916},
      archivePrefix={arXiv},
      primaryClass={cs.CL},
      url={https://arxiv.org/abs/2205.11916}, 
}

@misc{gpt4o,
      title={GPT-4o System Card}, 
    
      year={2024},
      eprint={2410.21276},
      archivePrefix={arXiv},
      primaryClass={cs.CL},
      url={https://arxiv.org/abs/2410.21276}, 
}

@article{Cyclomatic,
  title={Cyclomatic complexity},
  author={Ebert, Christof and Cain, James and Antoniol, Giuliano and Counsell, Steve and Laplante, Phillip},
  journal={IEEE software},
  volume={33},
  number={6},
  pages={27--29},
  year={2016},
  publisher={IEEE}
}

@article{pep8,
  title={PEP 8--style guide for python code},
  author={Van Rossum, Guido and Warsaw, Barry and Coghlan, Nick},
  journal={Python. org},
  volume={1565},
  pages={28},
  year={2001}
}

@article{jimenez2023swe-bench,
  title={Swe-bench: Can language models resolve real-world github issues?},
  author={Jimenez, Carlos E and Yang, John and Wettig, Alexander and Yao, Shunyu and Pei, Kexin and Press, Ofir and Narasimhan, Karthik},
  journal={arXiv preprint arXiv:2310.06770},
  year={2023}
}

\end{document}